\documentclass[10pt,a4paper]{article}
\usepackage[dvips]{color}
\usepackage{epsfig}
\usepackage{amsmath}
\usepackage{graphicx}

\begin{document}
\title{\bf Phenomenological Varying Modified Chaplygin Gas with Variable $G$ and $\Lambda$:\\ Toy Models for Our Universe}
\author{{J. Sadeghi$^{a}$ \thanks{Email: pouriya@ipm.ir},\hspace{1mm} M. Khurshudyan$^{b, c, d}$ \thanks{Email: martiros.khurshudyan@nano.cnr.it},\hspace{1mm}
and H. Farahani$^{a}$ \thanks{Email:
h.farahani@umz.ac.ir}}\\
$^{a}${\small {\em Department of Physics, Mazandaran University, Babolsar, Iran}}\\
{\small {\em P .O .Box 47416-95447, Babolsar, Iran}}\\
$^{b}${\small {\em CNR NANO Research Center S3, Via Campi 213a, 41125 Modena MO}}\\
$^{c}${\small {\em Dipartimento di Scienze Fisiche, Informatiche e Matematiche,}}\\
{\small {\em Universita degli Studi di Modena e Reggio Emilia, Modena, Italy}}\\
$^{d}${\small {\em Department of Theoretical Physics, Yerevan State
University, 1 Alex Manookian, 0025, Yerevan, Armenia}}}  \maketitle

\begin{abstract}
This article motivated by the recent articles and results of two
authors. Recently,  J. Sadeghi and H. Farahani presented a work [1],
where they include viscosity and analyze general model, by this way
they extended models considered by M. Khurshudyan [2] and [3]. In
this article, We tempt to consider varying Modified Chaplygin gas
model in case of variable $G$ and $\Lambda$. It is well known, that
varying $G$ and $\Lambda$ gives rise to modified field equations and
modified conservation laws. We will consider two different toy
models. First model is a Universe with one component
phenomenological gas of our consideration, while for the second
model we assume existence of a composed fluid of gas and a matter
with $P=\omega(t)\rho_{m}$. Sign changeable interaction between
fluids is accepted. We will analyze important cosmological
parameters like EoS parameter of a fluid, deceleration parameter $q$
of the model.
\end{abstract}
\section*{\large{Introduction}}
The observations of high redshift type SNIa supernovae [4-6] reveal
the speeding up expansion of our universe. The surveys of clusters
of galaxies show that the density of matter is very much less than
critical density [7], observations of Cosmic Microwave Background
(CMB) anisotropy indicate that the universe is flat and the total
energy density is very close to the critical $\Omega_{\small{tot}}
\simeq1$ [8]. In order to explain experimental data concerning to
the nature of the accelerated expansion of the Universe a huge
number of hypothesis were proposed. For instance, in general
relativity framework, the desirable result could be achieved by
so-called dark energy: an exotic and mysterious component of the
Universe, with negative pressure (we thought that the energy density
is always positive) and with negative EoS parameter $\omega<0$. Dark
energy occupies about 73$\% $ of the energy of our universe, other
component, dark matter, about 23$\%$, and usual baryonic matter
occupy about 4$\%$. The simplest model for a dark energy is a
cosmological constant $\omega_{\Lambda}=-1$ introduced by Einstein,
but with cosmological constant we faced with two problems i.e.
absence of a fundamental mechanism which sets the cosmological
constant zero or very small value the problem known as fine-tuning
problem, because in the framework of quantum field theory, the
expectation value of vacuum energy is 123 order of magnitude larger
than the observed value [9]. The second problem known as
cosmological coincidence problem, which asks why are we living in an
epoch in which the densities of dark energy and matter are
comparable? Alternative models of dark energy suggest a dynamical
form of dark energy, which at least in an effective level, can
originate from a variable cosmological constant [10, 11], or from
various fields, such as a canonical scalar field [12-17]
(quintessence), a phantom field, that is a scalar field with a
negative sign of the kinetic term [18-26] or the combination of
quintessence and phantom in a unified model named quintom [27-40]
and could alleviate these problems. Finally, an interesting attempt
to probe the nature of dark energy according to some basic quantum
gravitational principles are the holographic dark energy paradigm
[41-52] and agegraphic dark energy models [53-55].\\
Furthermore, since no known symmetry in nature prevents or
suppresses a non-minimal coupling between dark energy and dark
matter, there may exist interactions between the two components. At
the same time, from observation side, no piece of evidence has been
so far presented against such interactions. Indeed, possible
interactions between the two dark components have been discussed in
recent years. It is found that a suitable interaction can help to
alleviate the coincidence problem. Different interacting models of
dark energy have been investigated. For instance, the interacting
Chaplygin gas allows the universe to cross the phantom divide, which
is not permissible in pure Chaplygin gas models. This is a recent
believe and deep accepted fact among researchers. Several different
approaches are taken and intensively developed during these years in
order to find an appropriate physics. For some cases it can be
thought that not only one certain type of physics is involved and
stands behind this phenomena, but rather we are working with a
mixture of several physics giving the recent picture of our
Universe. There is deep hope concerning to the development of basis
of quantum gravity and respectively quantum cosmology. For any case
we faced with very hard problem. Despite to the huge number of
theoretical models and observational data and possibilities, still
we are working in darkness and we can not find that small light: the
end of tunnel. Research in theoretical cosmology considered two
possible ways to explain later time accelerated expansion of the
universe. Remembering the field equations, it becomes clear why.
Remember that field equations make connection between geometry and
matter content of Universe in a simple way. Therefore there is two
possibilities either we should modify matter content which is coded
in energy-stress tensor or we should modify geometrical part
including different functions of Ricci scalar etc. Different type of
couplings between geometry and matter could give desirable effects
as well.\\
Let give some critics concerning to mentioned ways thought to be
true. Mater content modification gives possibility to include two
types of matter: Dark energy (DE) and dark matter (DM). DE is a
fluid with negative pressure and positive energy density (remember
this is in classical regime, with no well established quantum
gravity coming from investigations of black holes Stephan Howking
accepted possibility of existence of dark energy with negative
energy density and positive pressure) giving negative EoS parameter
defined as $\omega=p/\rho$. Today, we do not feel lack of models for
DE, which could be seen from the number of references given above.
However, all of them are phenomenological models and wait to be
proved by observational data. The same can be said for DM, which
thought to operate on large scales and be responsible for structure
formation, evolution etc. There is a thought that these two
components are the manifestation of the same "matter". Even if it is
true, then there is not any mechanism explaining how it was or it is
possible to have such type of separation. Then, if you can accept
this possibility, then natural question is what will happen in
future i.e. can they recombine together or how they will start to
evolve and what are consequences. As a result of imagination, we can
propose several mechanism that can make connection between DE and DM
and one of them is possibility having speeds faster than speed of
light, making connection mysterious for our brain, because simply we
can not "see" that. But if they have different origin, then we
really face with a wall. However, independent of the mentioned and
other possibilities, we have big open problem, because we do not
know structure of these matters. Research concerning to the
modification matter part of field equations give rise of an
understanding that more complex and crazy forms for EoS equation
can be considered.\\
Modification of geometrical part gives rise of different modified
theories like $F(R)$, $F(T)$, $F(G)$ etc. Models of this origin
however contain some future singularities that our Universe should
faced, but we do not need to worry, because these are just models
and fortunately with appropriate choose of a function we can escape
singularities and extend life-time for our Universe. But, there are
models also, that can explain accelerated expansion without any DE,
for instance, Cardassian Universe [56-61]. In this model, we need
modify Friedmann equations and having usual matter is enough.\\
It is well known that Einstein equations of general relativity do
not permit any variation in the gravitational constant $G$ and
cosmological constant $\Lambda$ because of the fact that the
Einstein tensor has zero divergence and by energy conservation law
is also zero. So, some modifications of Einstein equations are
necessary. This is because, if we simply allow $G$ and $\Lambda$ to
be a variable in Einstein equations, then energy conservation law is
violated. Therefore, the study of the effect of varying $G$ and
$\Lambda$ can be done only through modified field equations and
modified conservation laws.\\
In this paper we use results of the resent papers [1-3] and extend
them to the case of variable $G$ and $\Lambda$. We will consider two
different toy models. First model is a Universe with one component
phenomenological gas of our consideration, while for the second
model we assume existence of a composed fluid of gas and a matter.
We use sign changeable interaction between fluids and analyze
important cosmological parameters like EoS parameter of a fluid and
deceleration parameter $q$ of the model. This article is organized
in following way. Section introduction is devoted to introduce basic
ideas and gives some general information related to the research
field and our motivation. Next section review FRW Universe with
variable $G$ and $\Lambda$. In section "Phenomenological Fluid and
Model Setup" we recall the basics of origin of the fluid and general
settings how problem can be solved. In other sections we investigate
parameters of the models and then we give conclusions.

\section*{\large{FRW Universe with variable $G$ and $\Lambda$}}
Field equations that govern our model with variable $G(t)$ and
$\Lambda(t)$ (see for instance [62]) are,
\begin{equation}\label{s1}
R^{ij}-\frac{1}{2}Rg^{ij}=-8 \pi G(t) \left[ T^{ij} -
\frac{\Lambda(t)}{8 \pi G(t)}g^{ij} \right],
\end{equation}
where $G(t)$ and $\Lambda(t)$ are function of time. By using the
following FRW metric for a flat Universe,
\begin{equation}\label{s2}
ds^2=-dt^2+a(t)^2\left(dr^{2}+r^{2}d\Omega^{2}\right),
\end{equation}
field equations can be reduced to the following Friedmann equations,
\begin{equation}\label{eq: Fridmman vlambda}
H^{2}=\frac{\dot{a}^{2}}{a^{2}}=\frac{8\pi G(t)\rho}{3}+\frac{\Lambda(t)}{3},
\end{equation}
and,
\begin{equation}\label{eq:fridman2}
\frac{\ddot{a}}{a}=-\frac{4\pi
G(t)}{3}(\rho+3P)+\frac{\Lambda(t)}{3},
\end{equation}
where $d\Omega^{2}=d\theta^{2}+\sin^{2}\theta d\phi^{2}$, and $a(t)$
represents the scale factor. The $\theta$ and $\phi$ parameters are
the usual azimuthal and polar angles of spherical coordinates, with
$0\leq\theta\leq\pi$ and $0\leq\phi<2\pi$. The coordinates ($t, r,
\theta, \phi$) are called co-moving coordinates.\\
Energy conservation $T^{;j}_{ij}=0$ reads as,
\begin{equation}\label{eq:conservation}
\dot{\rho}+3H(\rho+P)=0.
\end{equation}
Combination of (3), (4) and (5) gives the relationship between
$\dot{G}(t)$ and $\dot{\Lambda}(t)$
\begin{equation}\label{eq:glambda}
\dot{G}=-\frac{\dot{\Lambda}}{8\pi\rho}.
\end{equation}
Ever since Dirac's proposition of a possible time variation of $G$,
a volume of works has been centered around the act of calculating
the amount of variation of the gravitational constant. For instance,
observation of spinning-down rate of pulsar $PSR J2019+2425$
provides the result,
\begin{equation}\label{eq:gvar1}
\left|\frac{\dot{G}}{G} \right|\leq (1.4-3.2) \times 10^{-11} yr^{-1}.
\end{equation}
Depending on the observations of pulsating white dwarf star $G 117-B
15A$, Benvenuto et al. [63] have set up the astroseismological bound
as,
\begin{equation}\label{eq:gvar2}
 -2.50 \times 10^{-10} \leq \left|\frac{\dot{G}}{G} \right|\leq 4 \times 10^{-10} yr^{-1}.
\end{equation}
For a review to "Large Number Hypothesis" (LNH) we refer our readers
to
\cite{Saibal} and references therein.\\
Subject of our interest is to consider a modified Chaplygin gas
based on a Universe with variable $G$ and $\Lambda$. This model is a
phenomenological and we are interested by the evolution of the
Universe with this setup. For $\Lambda$ we will consider one of the
forms  intensively considered in literature and using Eq.s (3)-(6)
we can recover scale factor, behavior of $G$ and other cosmological
parameters including EoS parameter and deceleration parameter $q$,
which could describe behavior of the Universe. The starting point
for scanning the range of parameters of the model is that
acceleration will be caused by a fluid with negative EoS parameter
i.e. at the same time we should satisfy $q<0$ and $\omega <0$
conditions. Keeping spirit of Dirac's LNH we will consider,
\begin{equation}\label{eq:vlambda}
\Lambda \propto t^{-2}.
\end{equation}
Other forms for $\Lambda(t)$ were considered over years based on
phenomenological approach, some of examples are, for instance,
$\Lambda \propto (\dot{a}/a)$, $\Lambda \propto \ddot{a}/a$ or
$\Lambda \propto \rho$ to mention a few. As we are interested by toy
models we pay our attention to the problem from a numerical
investigation point of view and we believe that after some effort we
also can provide exact solutions for the problem, which will be done
in other forthcoming articles.

\section*{\large{Phenomenological Fluid and Model Setups}}
We already mentioned, that matter modification of field equations
teach us that we can consider fluids with strange EoS equations. One
of the examples is Chaplygin gas given by [65],
\begin{equation}\label{eq:chgas}
P=-\frac{B}{\rho},
\end{equation}
where $B$ is a constant. Then, generalized Chaplygin gas (GCG) with
the following equation of state [66, 67],
\begin{equation}\label{eq:gcg}
P=-\frac{B}{\rho^{\alpha}},
\end{equation}
with $0<\alpha \leq 1$. As we can see the GCG is corresponding to
almost dust ($P = 0$) at high density which is not agree completely
with our Universe. Therefore, modified Chaplygin gas (MCG) with the
following equation of state introduced [68-70],
\begin{equation}\label{eq:mcg}
P=\mu \rho -\frac{B}{\rho^{\alpha}},
\end{equation}
where $\mu$ is a positive constant. This model is more appropriate
choice to have constant negative pressure at low energy density and
high pressure at high energy density. The special case of
$\mu=\frac{1}{3}$ is the best fitted value to describe evolution of
the universe from radiation regime to the $\Lambda$CDM regime.  In
the Ref. [2] one of the authors motivated by a series of works
[71-73] proposed a model of varying generalized Chaplygin gas and
considered its sign-changeable interaction of the form
$Q=q(3Hb\rho+\gamma \dot{\rho} )$ with fluid. As there is an
interaction between components there is not energy conservation for
the components separately, but for the whole mixture the energy
conservation is hold. This approach could work as long as we are
working without knowing the actual nature of the dark energy and
dark matter as well as about the nature of the interaction. This
approach at least from mathematical point of view is correct. The
forms of interaction term considered in literature very often are of
the following forms: $Q=3Hb\rho_{dm}$, $Q=3Hb\rho_{\small{de}}$,
$Q=3Hb\rho_{\small{tot}}$, where $b$ is a coupling constant and
positive $b$ means that dark energy decays into dark matter, while
negative $b$ means dark matter decays into dark energy. From
thermodynamical view, it is argued that the second law of
thermodynamics strongly favors dark energy decays into dark matter.
However it was found that the observations may favor the decaying of
dark matter into dark energy. Other forms for interaction term
considered in literature are $Q=\gamma\dot{\rho}_{\small{dm}}$,
$Q=\gamma\dot{\rho}_{\small{de}}$,
$Q=\gamma\dot{\rho}_{\small{tot}}$,
$Q=3Hb\gamma\rho_{i}+\gamma\dot{\rho_{i}}$, where $i=\{dm,de,tot\}$.
These type of interactions are either positive or negative and can
not change sign. However, recently by using a model independent
method to deal with the observational data Cai and Su found that the
sign of interaction $Q$ in the dark sector changed in the redshift
range of $0.45 \leq z \leq 0.9$.  Hereafter, a sign-changeable
interaction [74] and [75] were introduced,
\begin{equation}\label{eq:signcinteraction}
Q=q(\gamma\dot{\rho}+3b H\rho),
\end{equation}
where $\gamma$ and $b$ are dimensionless constants, the energy
density $\rho$ could be $\rho_{m}$, $\rho_{\small{de}}$,
$\rho_{tot}$. $q$ is the deceleration parameter given by,
\begin{equation}\label{eq:decparameter}
q=-\frac{1}{H^{2}} \frac{\ddot{a}}{a}=-1-\frac{\dot{H}}{H^{2}}.
\end{equation}
This new type of interaction, where deceleration parameter $q$ is a
key ingredient makes this type of interactions different from the
ones considered in literature and presented above, because it can
change its sign when our universe changes from deceleration $q>0$ to
acceleration $q<0$. $\gamma \dot{\rho}$ is introduced from the
dimensional point of view. We would like also to stress a fact, that
by this way we import a more information about the geometry of the
Universe into the interaction term. This is not only one
possibility, in forthcoming articles we hope to provide and consider
other forms of sign-changeable interactions. The origin of the fluid
considered before, is based on a simple assumption. We proposed the
following strategy: somehow, concerning to an unknown physics (even
could be very well known) it was possible to separate components of
the darkness of the Universe and was found that we have fluid (F) +
MCG + Remaining Darkness (RD). Only information which we know about
RD is that its EoS parameter $\omega_{x} < 0$. We assume that
interaction between MCG and RD gives to born a varying modified
Chaplygin gas and that there is not any interaction between F and
RD. This assumption allows us still to think that our Universe
consists of mixture of a F and a varying MCG gas. Then we will
assume that EoS parameter of RD $\omega_{x}$ is a function of time:
$\omega(t) = \omega_{0}+\omega_{1}(t \frac{\dot{H}}{H})$, which has
an explicit time dependence that disappears with the $t \dot{H}=H$
condition. Which in its turn gives us a modification of $B$ constant
and for the new model it reads as,
\begin{equation}\label{eq:Bvarying}
B(a)= -\omega(t)B_{0}a^{-3(1+\omega(t))(1+\alpha)}.
\end{equation}
Therefore a fluid of our Universe considered to be a MCG with the
following EoS,
\begin{equation}\label{eq:CGfinal}
P=\mu \rho - \frac{B(a)}{\rho^{\alpha}}.
\end{equation}
In order to have a complete picture of our proposition, we will
consider two different models as the following.
\begin{enumerate}
\item An Universe with varying $G$ and $\Lambda$ with a given $\Lambda(t)$ and Varying MCG
\item An Universe with varying $G$ and $\Lambda$ with a given $\Lambda(t)$ and a mixture of Varying MCG and a fluid.
\end{enumerate}
For the fluid we assume a EoS $P_{f}=\omega(t)\rho_{f}$, where
$\omega(t)$ is the same as for varying MCG of our consideration.
Possible coupling between fluids is modeled by sigh-changeable
interaction.

\section*{\large{Models and Cosmological Parameters}}
In this section we will start to investigate models and we will
start with a single component Universe. Bellow we provide general
algorithm which a reader should keep in order to perform numerical
research of the problem. For simplicity, we reduce number of
parameters of the models assuming some of them known a priori.
\subsection*{One component fluid Universe}
Compared with the second model of our interest first model is going
to be a simple model. Problem solving strategy is based on a
solution of a system of differential equations (4), (5) and (6) with
a given $\Lambda(t) \propto t^{-2}$ in our case. We analyze problem
numerically and present profiles of cosmological parameters and
quantities and give comprehensive analysis of obtained results.
Plots are organized in such a way, that they provide good inside
into properties of the model.\\
First panel of $4$ plots (Fig. 1) represent behavior of $G(t)$ for
different values of model parameters. From the first graph
representing behavior of $G(t)$ as a function of $\omega_{0}$ with
$\omega_{1}=-1.8$ and $\alpha=0.3$ as follow: $G(t)$ is an
increasing function. For some initial stages of evolution $G(t)$ has
almost the same behavior, which starts exhibit differently for
latter stages evolution.  Behavior of $G(t)$ reveal to be the same
for a different values of $\alpha$ for a given values of
$\omega_{0}$ and $\omega_{1}$. From the third graph, we can
understood that with decreasing value of $\omega_{1}$, after certain
time we register decreasing in rate of $G(t)$. Last graph is
represent general case for the behavior of $G$.\\
Second panel of graphs (Fig. 2) represent behavior of deceleration
parameter $q$ over time. We have ever accelerated expansion, because
$q$ is negative during whole evolution. For different scenarios
related to pair of fixed and varying parameters $q$ increases for
early stages of evolution, but for later stages of evolution it is
decreases. However, with $\omega_{1}=-1.8$ and $\alpha=0.3$ we
observed that $q$ continues its decreasing with a law speed in case
$\omega_{0}=-2.2$. For a fixed values of $\omega_{0}$ and
$\omega_{1}$ for different values of $\alpha$ we conclude that we
have increasing and then decreasing function of $q$.\\
The same discussion hold for third panel of graphs (Fig. 3) which
represent $\omega_{MCG}$ in terms of time.\\

\begin{figure}[h]
 \begin{center}$
 \begin{array}{cccc}
\includegraphics[width=50 mm]{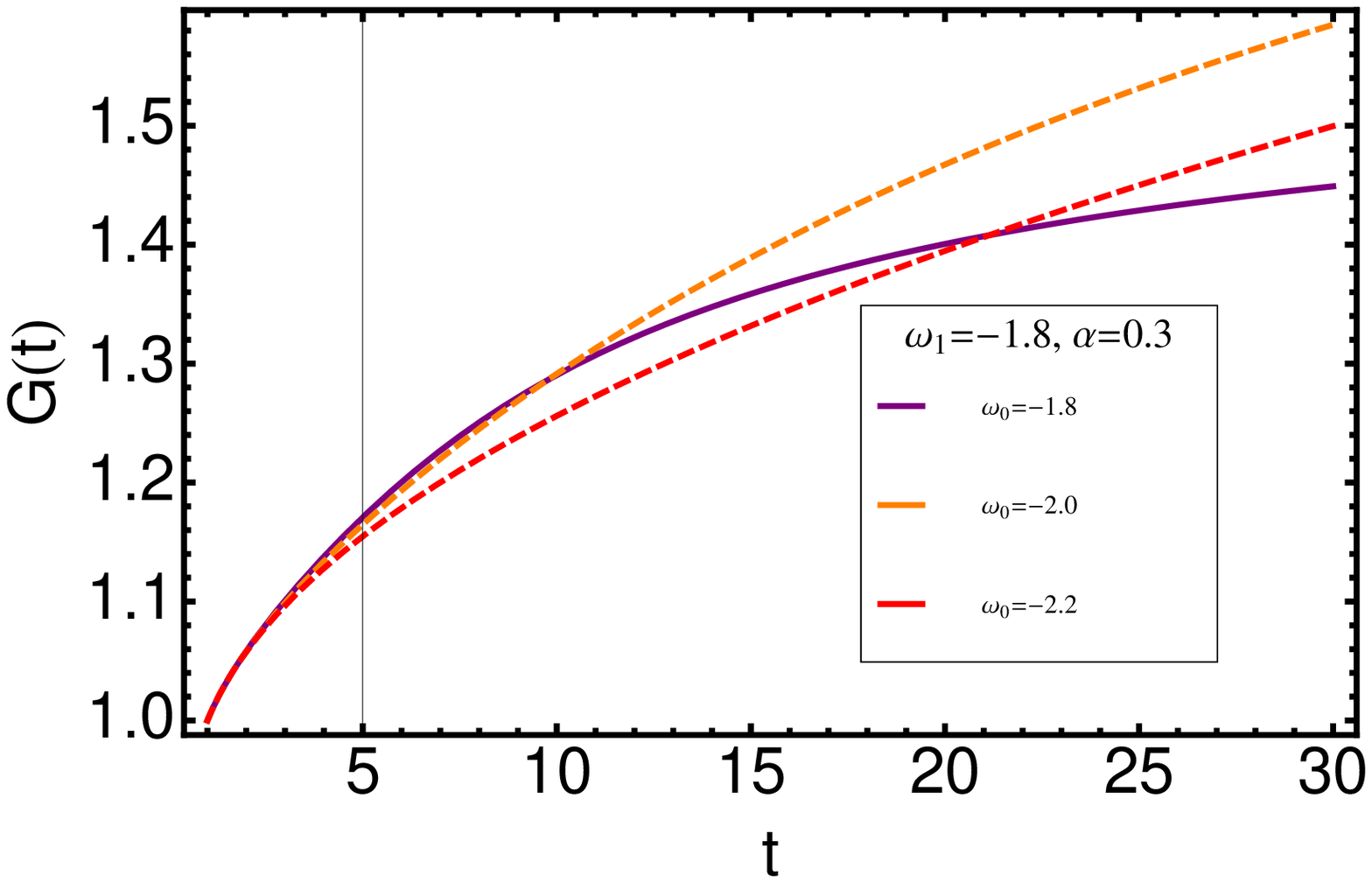} &
\includegraphics[width=50 mm]{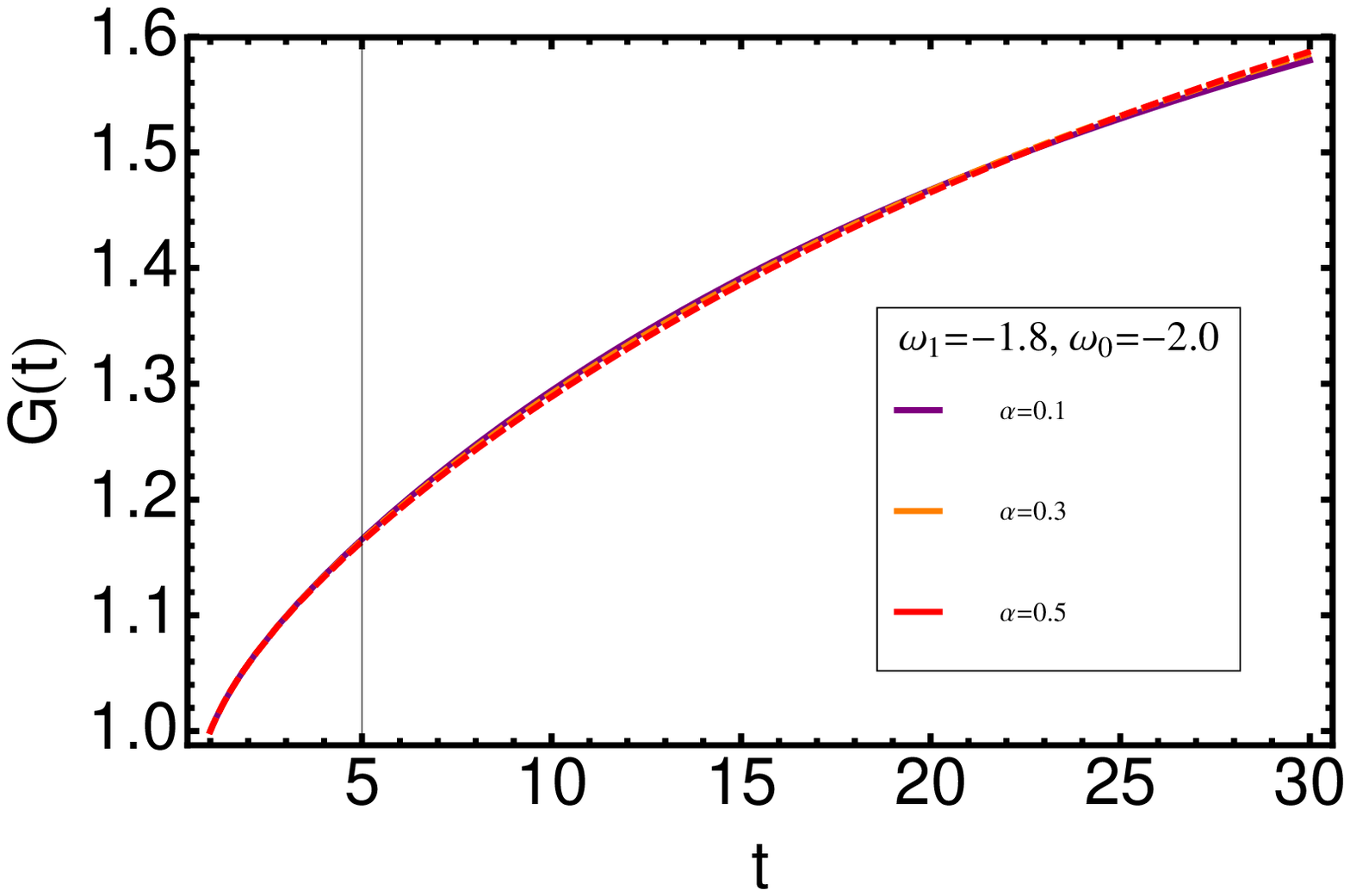}\\
\includegraphics[width=50 mm]{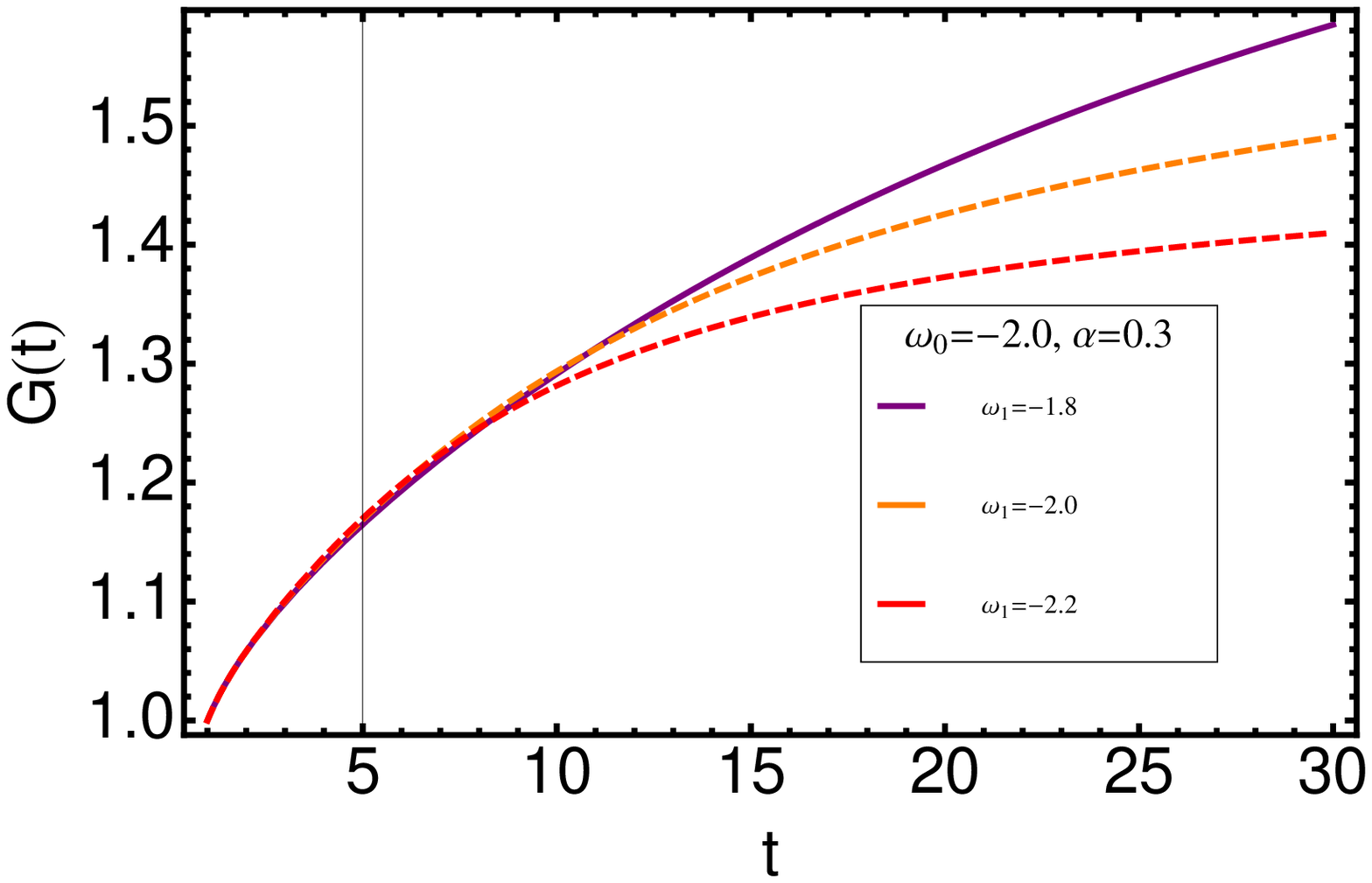} &
\includegraphics[width=50 mm]{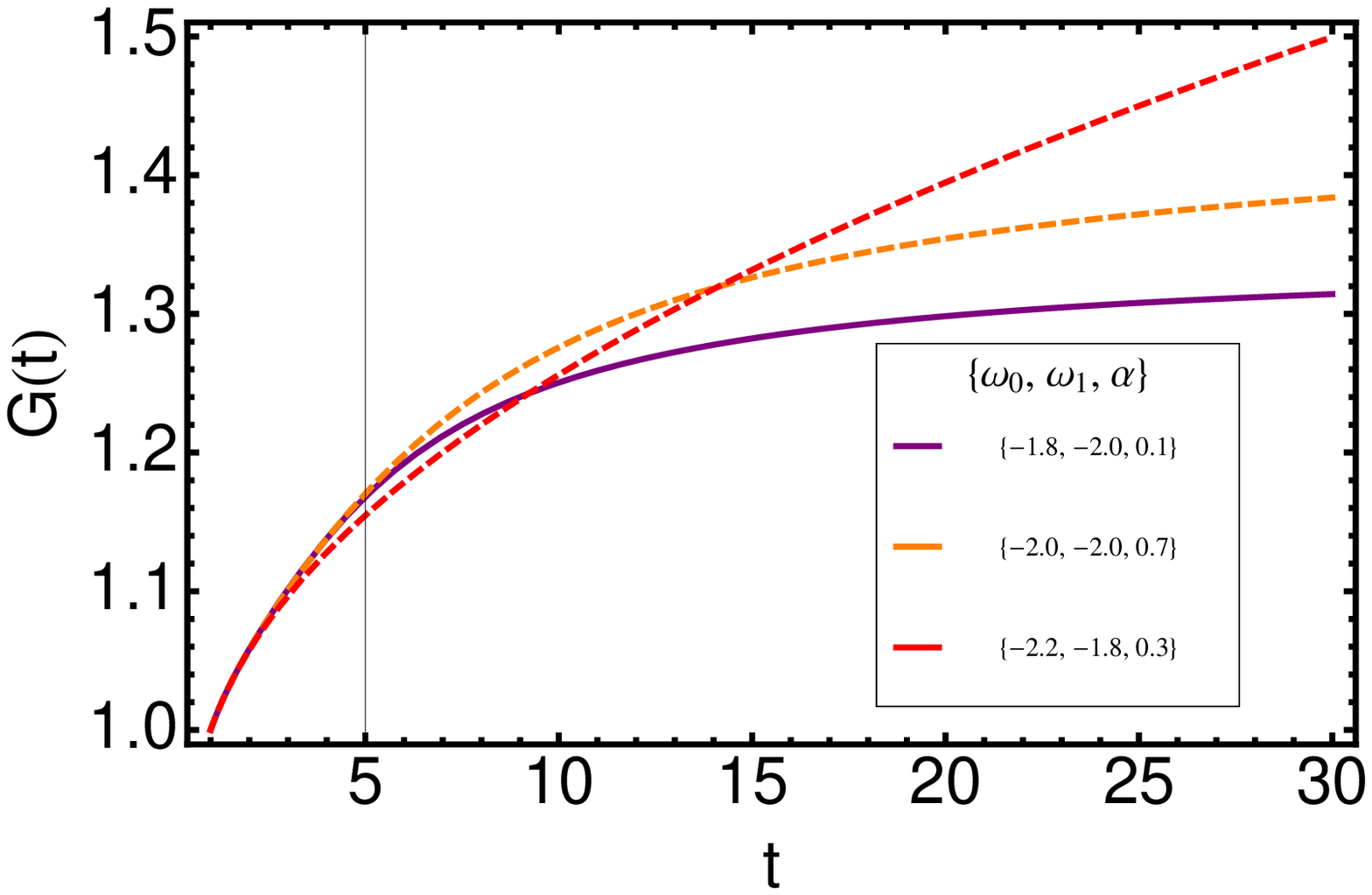}
 \end{array}$
 \end{center}
\caption{One component}
 \label{fig:1}
\end{figure}

\begin{figure}[h]
 \begin{center}$
 \begin{array}{cccc}
 \includegraphics[width=60 mm]{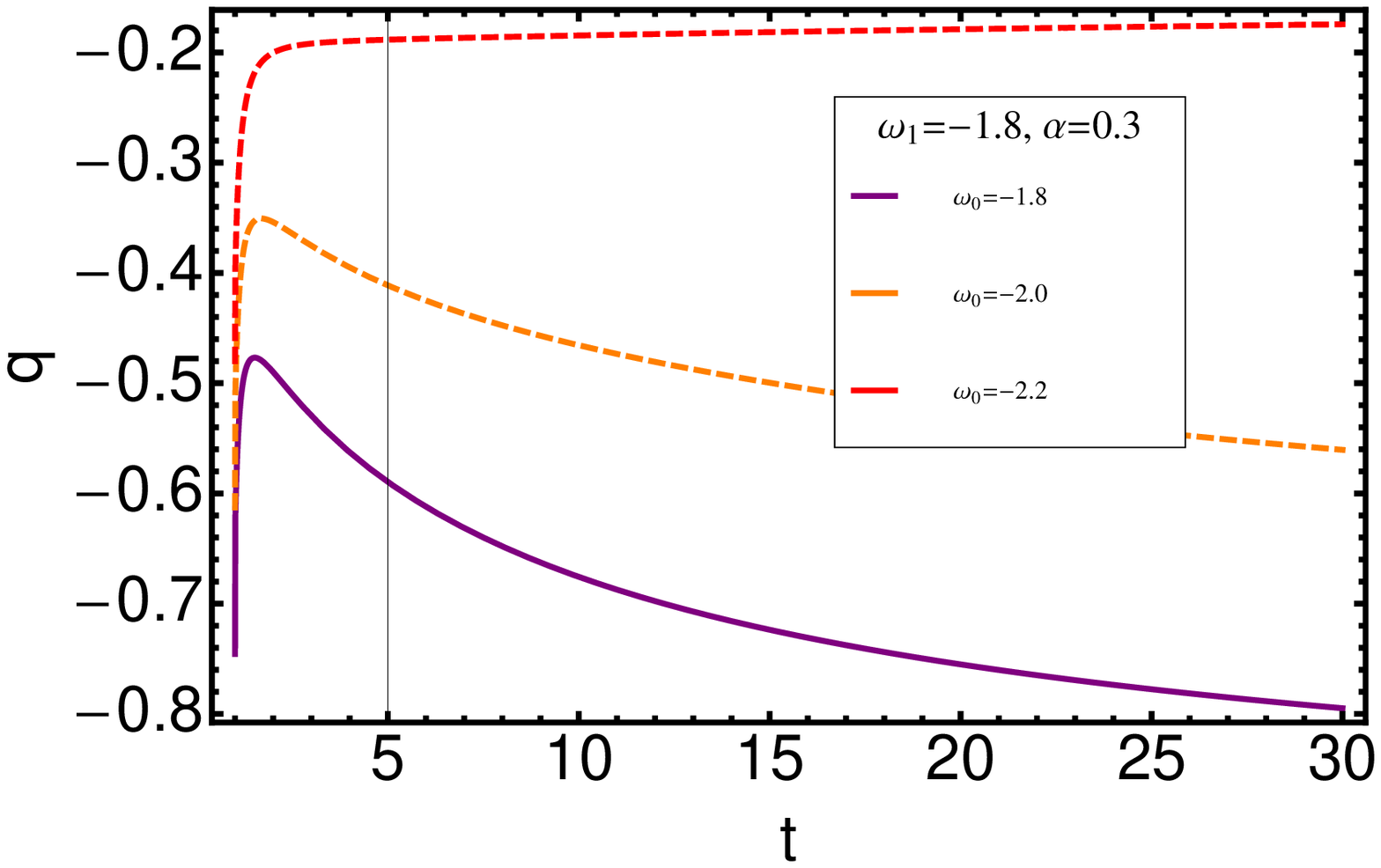} &
 \includegraphics[width=60 mm]{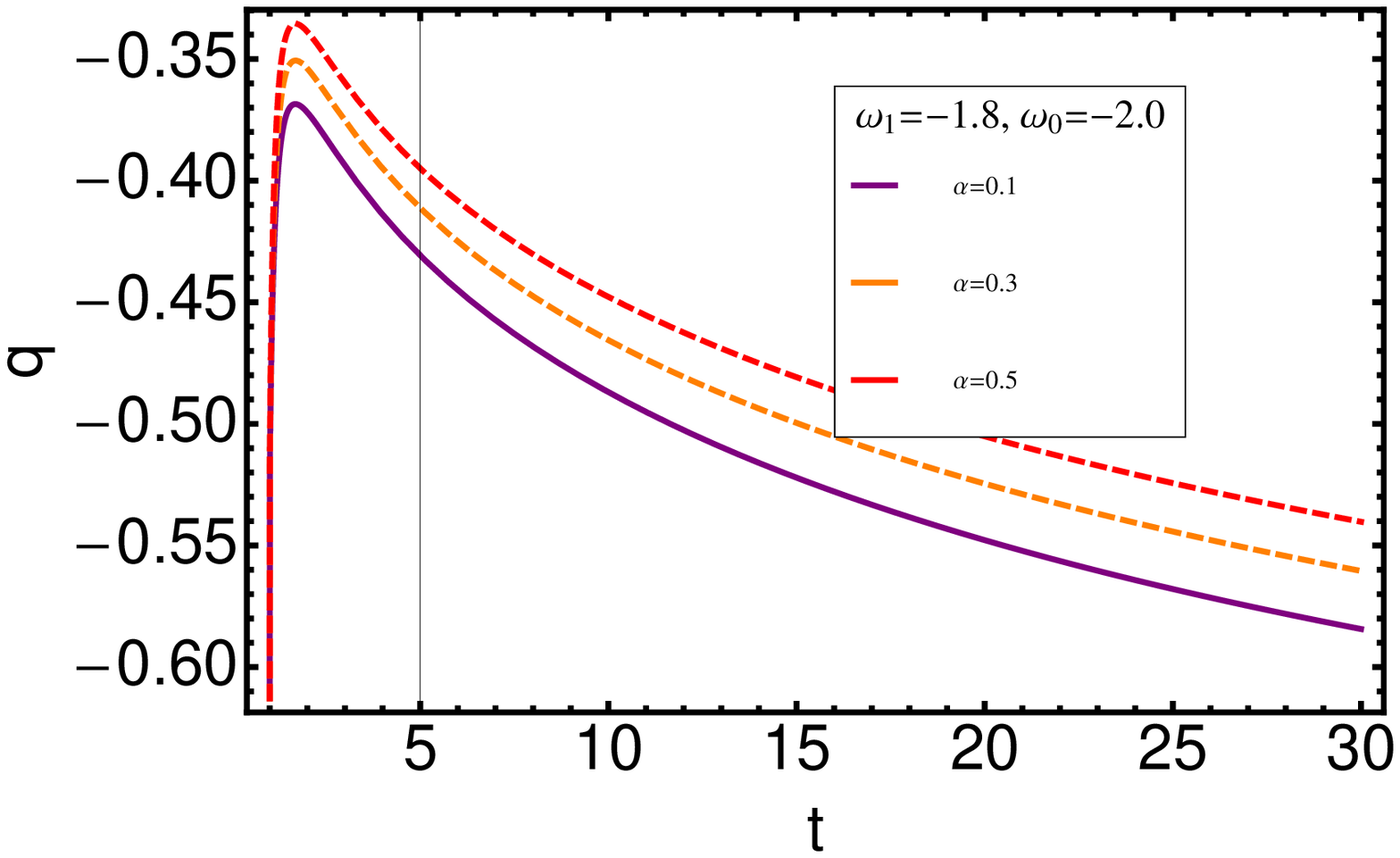} \\
 \includegraphics[width=60 mm]{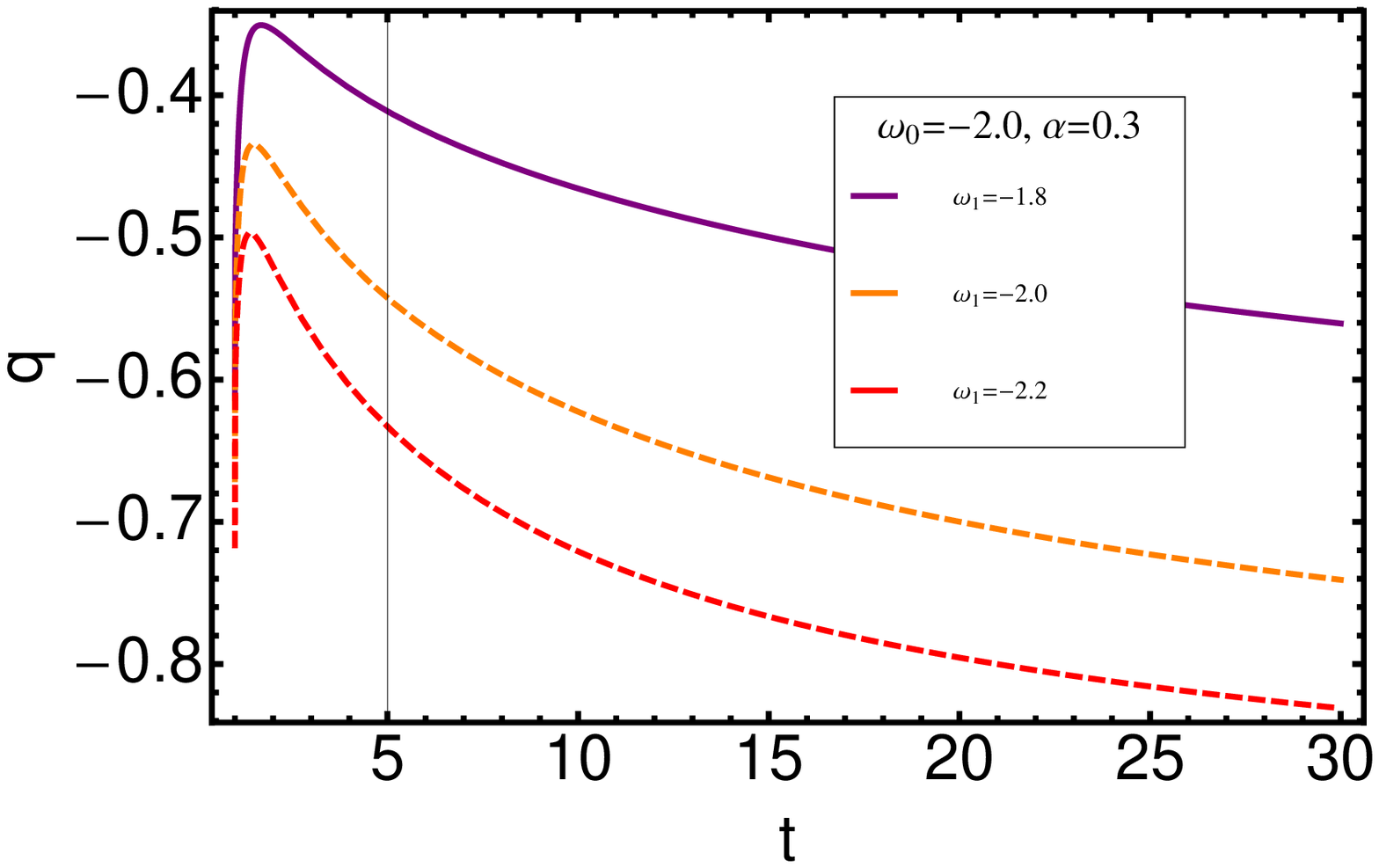} &
 \includegraphics[width=60 mm]{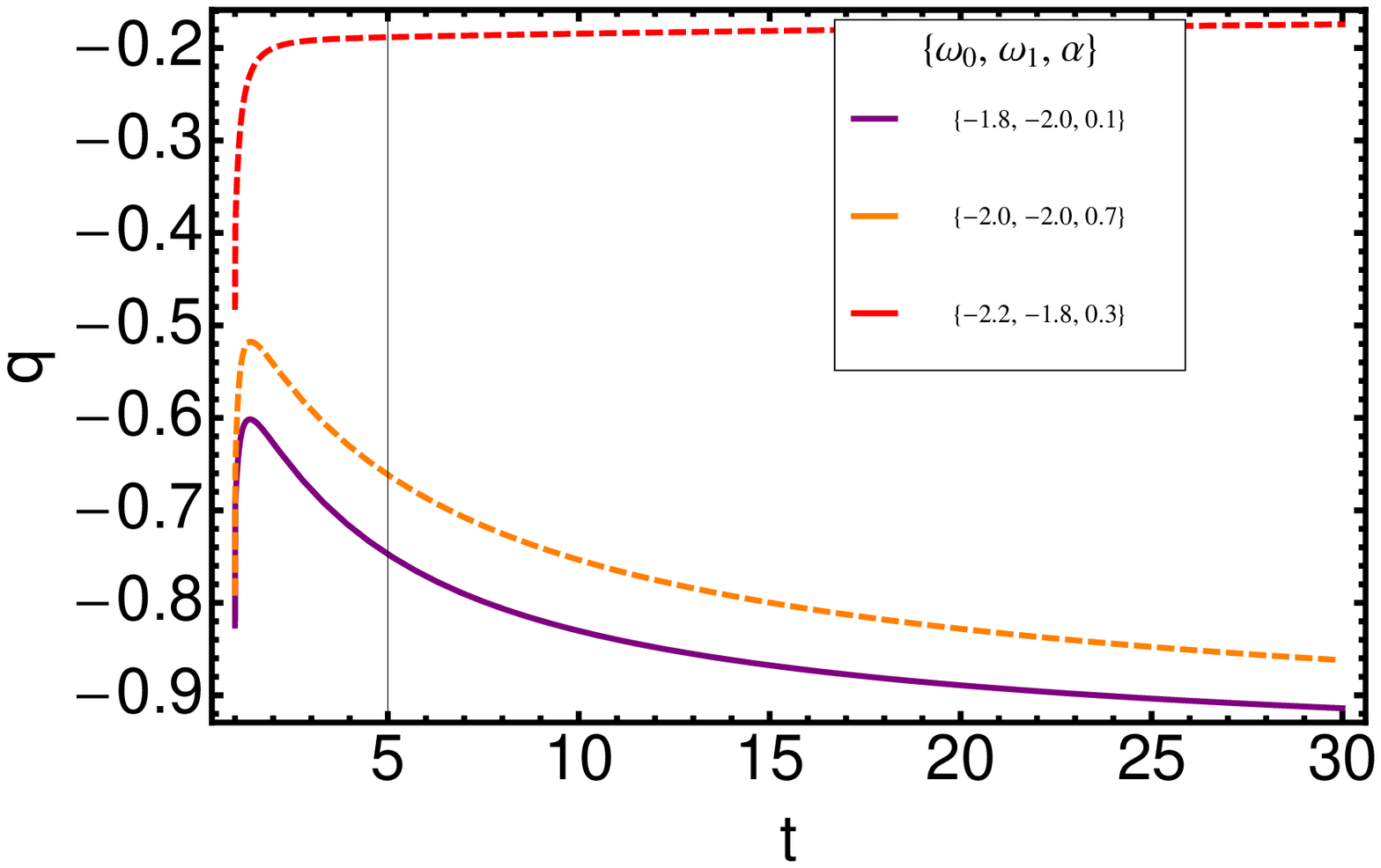}
 \end{array}$
 \end{center}
\caption{One component}
 \label{fig:2}
\end{figure}

\begin{figure}[h]
 \begin{center}$
 \begin{array}{cccc}
 \includegraphics[width=60 mm]{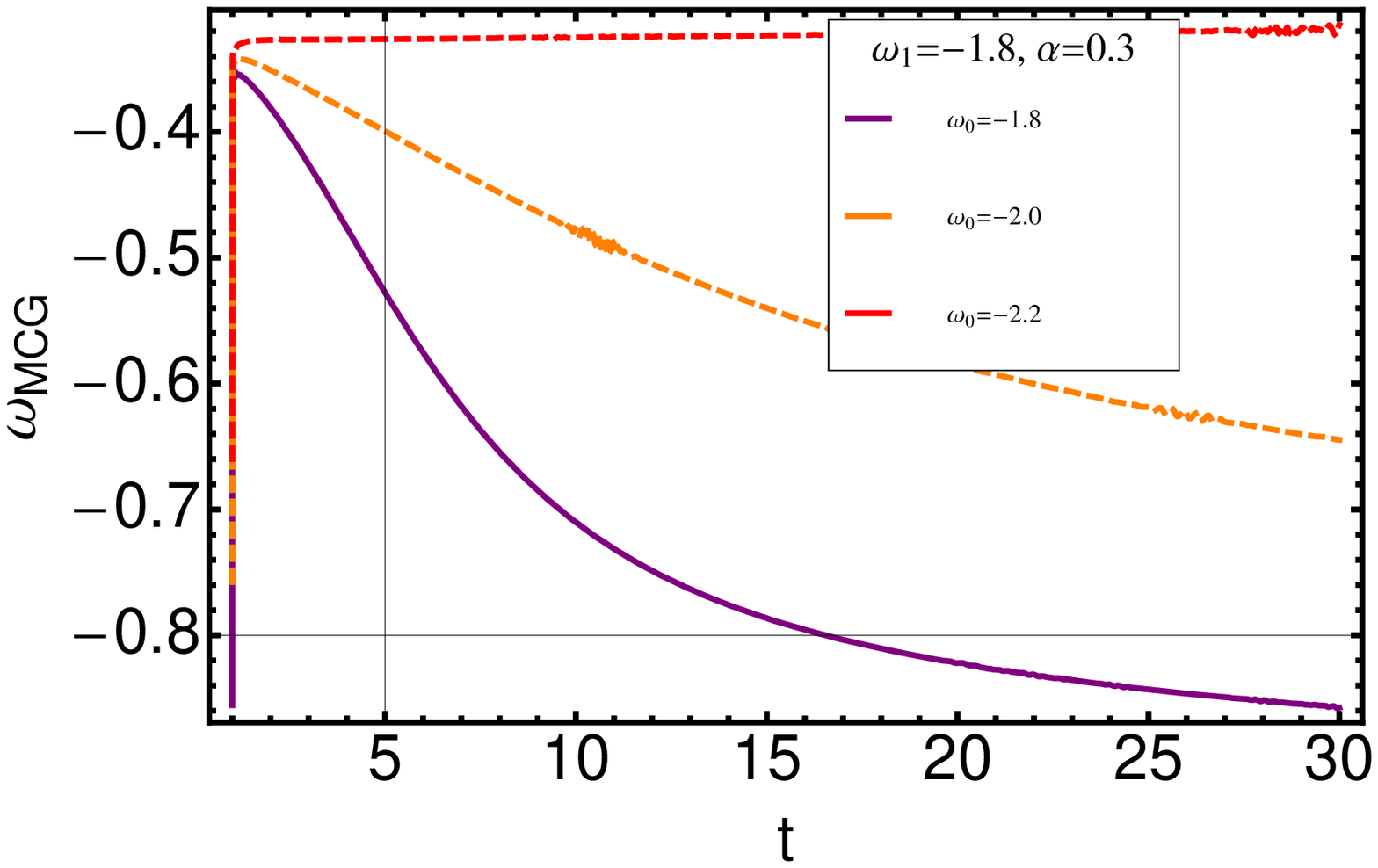} &
 \includegraphics[width=60 mm]{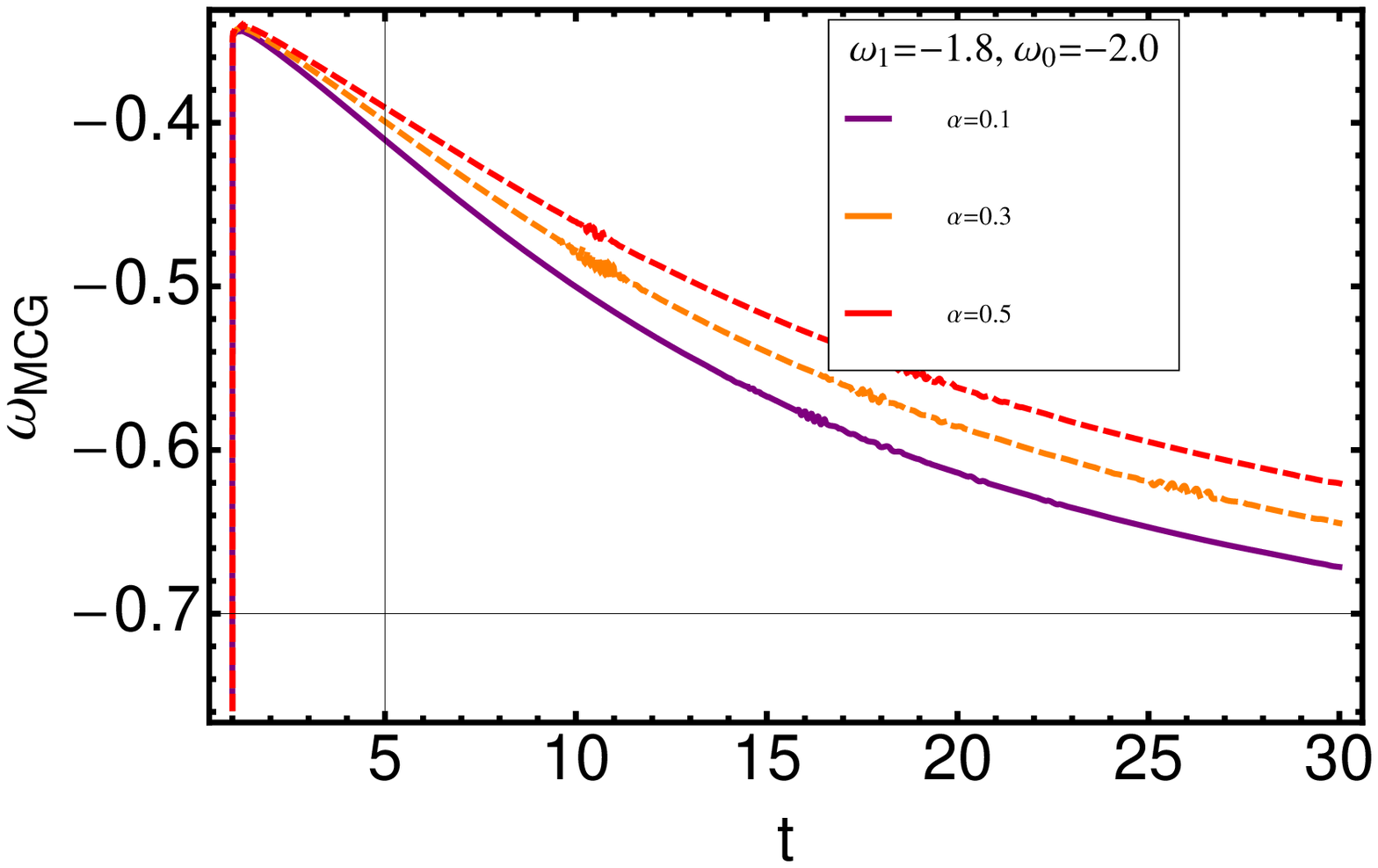} \\
 \includegraphics[width=60 mm]{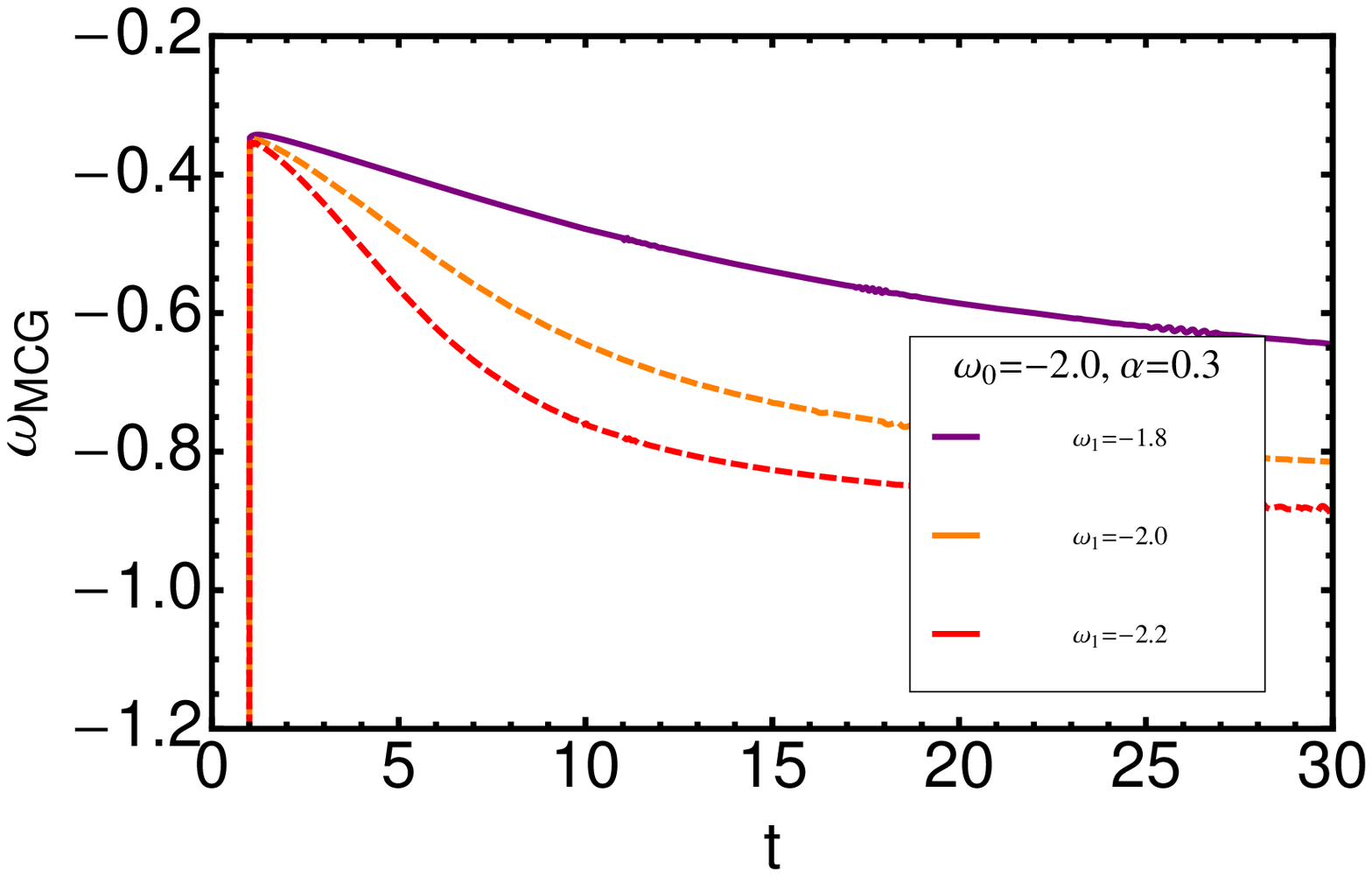} &
 \includegraphics[width=60 mm]{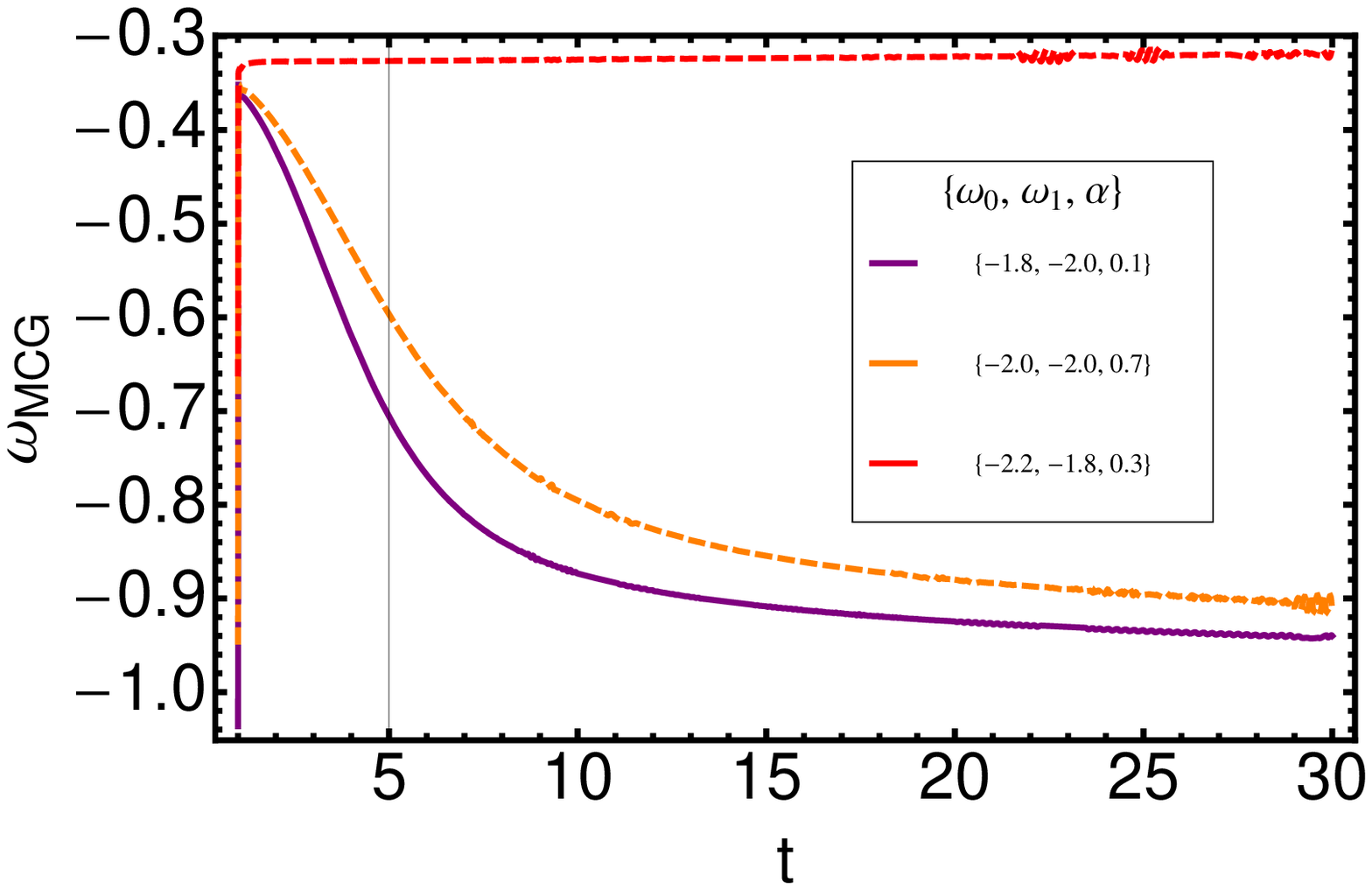}
 \end{array}$
 \end{center}
 \caption{One component}
 \label{fig:3}
\end{figure}

\subsection*{Two Component Fluid Universe}
If we consider two component fluid Universe, then equations (3) and
(4) extended to the following,
\begin{equation}\label{17}
(\frac{\dot{a}}{a})^{2}=\frac{8\pi
G(t)\rho_{tot}}{3}+\frac{C}{3t^{2}},
\end{equation}
where $C$ is an arbitrary constant and,
\begin{equation}\label{18}
\frac{\ddot{a}}{a}=-\frac{4\pi
G(t)}{3}(\rho_{tot}+3P_{tot})+\frac{C}{3t^{2}},
\end{equation}
where we used the equation (9) and define,
\begin{eqnarray}\label{19}
\rho_{tot}=\rho_{F}+\rho_{MCG},\nonumber\\
P_{tot}=P_{F}+P_{MCG}.
\end{eqnarray}
Therefore the energy-momentum conservation law obtained as the
following,
\begin{equation}\label{s20}
\dot{\rho}_{tot}+3(\rho_{tot}+{P_{tot}})H=0.
\end{equation}
Also we can obtain,
\begin{equation}\label{21}
\dot{G}=\frac{2C}{8\pi\rho_{tot}t^{3}}.
\end{equation}

\section*{\large{Interacting case}}
If we consider interaction between fluid and gas then the
conservation energy separate as the following,
\begin{equation}\label{s22}
\dot{\rho}_{MCG}+3H(\rho_{MCG}+{P_{MCG}})=Q,
\end{equation}
where $P_{MCG}$ is given by the relations (12) and,
\begin{equation}\label{s23}
\dot{\rho}_{F}+3H(\rho_{F}+{P_{F}})=-Q,
\end{equation}
where $Q$ is interaction term given by the equation (13). Now we
should solve a system of differential equations of (18), (21), (22)
and (23).\\
Plots given by the Fig. 4 represent behavior of $G(t)$ for different
values of model parameters. The first graph representing behavior of
$G(t)$ as a function of $\omega_{0}$ which shows increasing function
of time but yields to a constant at the late time which is
consequence of interaction term. Such behavior is the same for other
plots which are graphs representing behavior of $G(t)$ as a function
of $\alpha$, $\omega_{1}$ and $\gamma$ respectively. As $\alpha$ and
$\gamma$ increased then $G(t)$ increased, but increasing
$|\omega_{1}|$ decreased $G(t)$.\\
Plots of the Fig. 5 represent behavior of deceleration parameter $q$
over time. As previous case we have ever accelerated expansion,
because $q$ is negative during whole evolution. The last plot shows
that varying $\alpha$ is not important and $q$ is totally decreasing
function of time. Variation of other parameters show increases of
$q$ for early stages of evolution, but for later stages of evolution
it is decreases. The first plot shows that increasing $|\omega_{0}|$
increased the value of $q$ but increasing $|\omega_{1}|$ decreased
one.\\
Plots of the Fig. 6 represent $\omega_{MCG}$ versus time. We can see
that $\omega_{MCG}$ increased in the initial stage to reach maximum
at about $0.1<t<0.2$ and then decreased to reach a constant value at
the late time. This is completely different with the previous case
where there is only one component fluid. the second plot shows that
increasing $\alpha$ increased $\omega_{MCG}$. The last graph is
represent general case for the behavior of $\omega_{MCG}$.

\begin{figure}[h]
 \begin{center}$
 \begin{array}{cccc}
\includegraphics[width=50 mm]{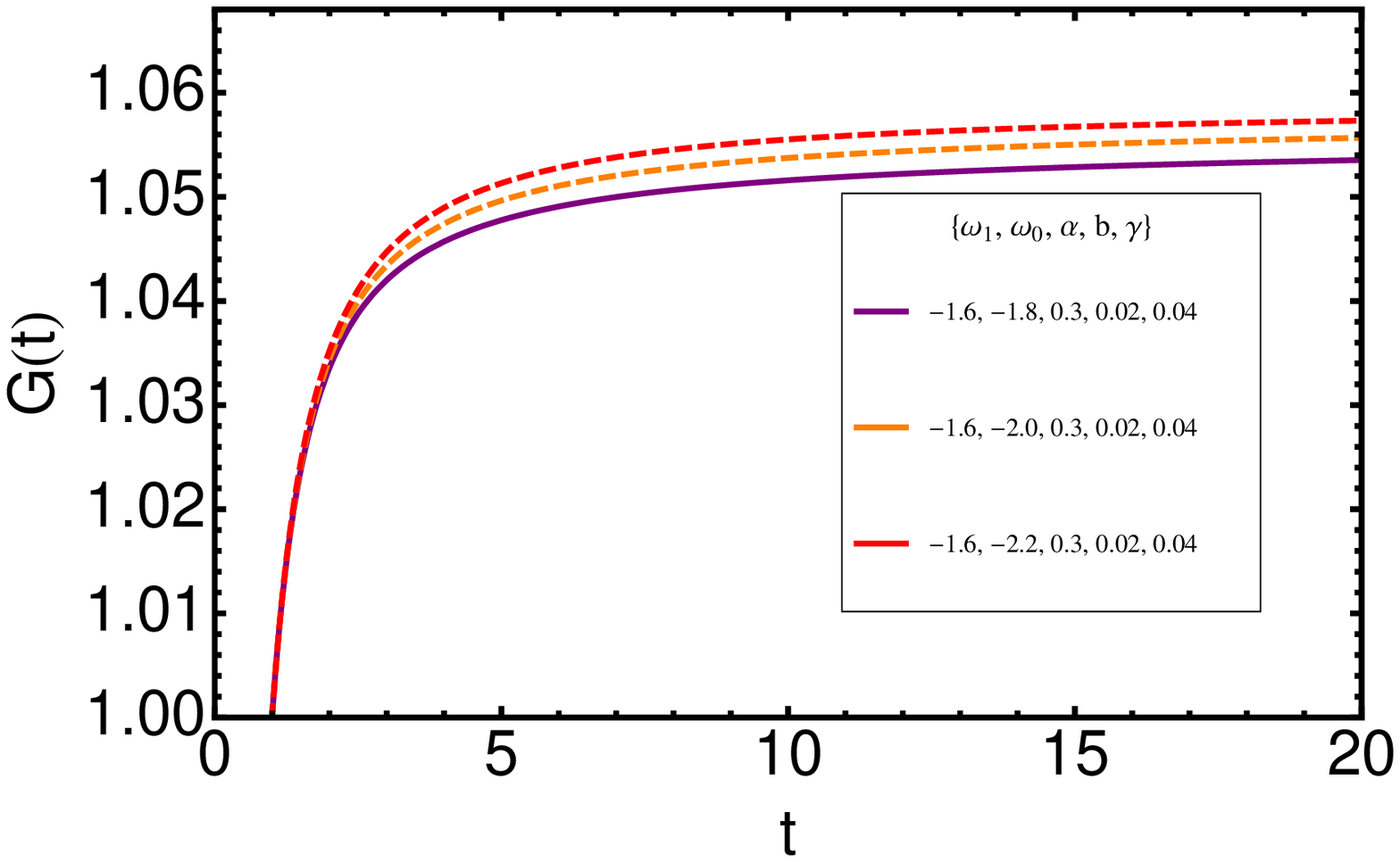} &
\includegraphics[width=50 mm]{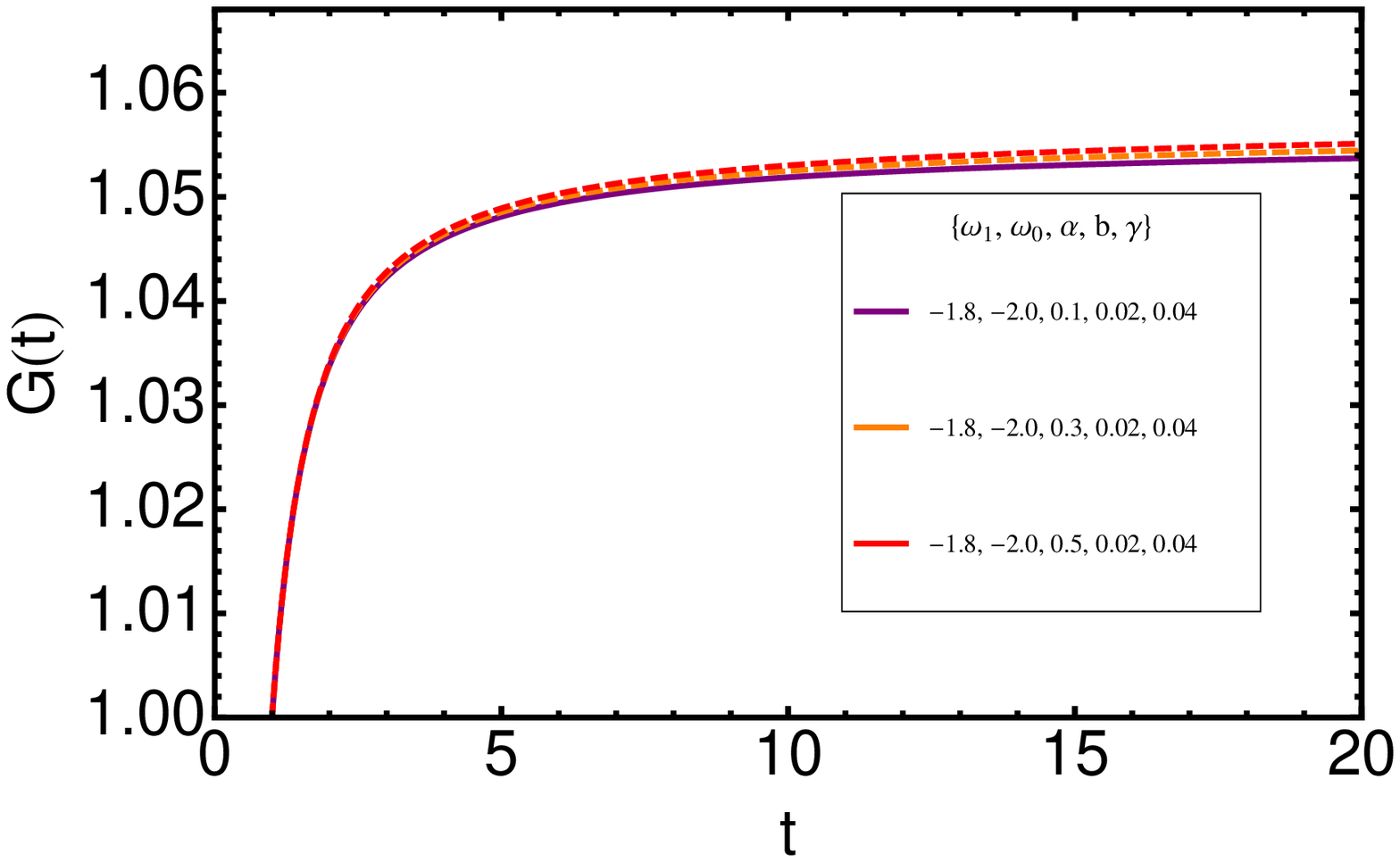}\\
\includegraphics[width=50 mm]{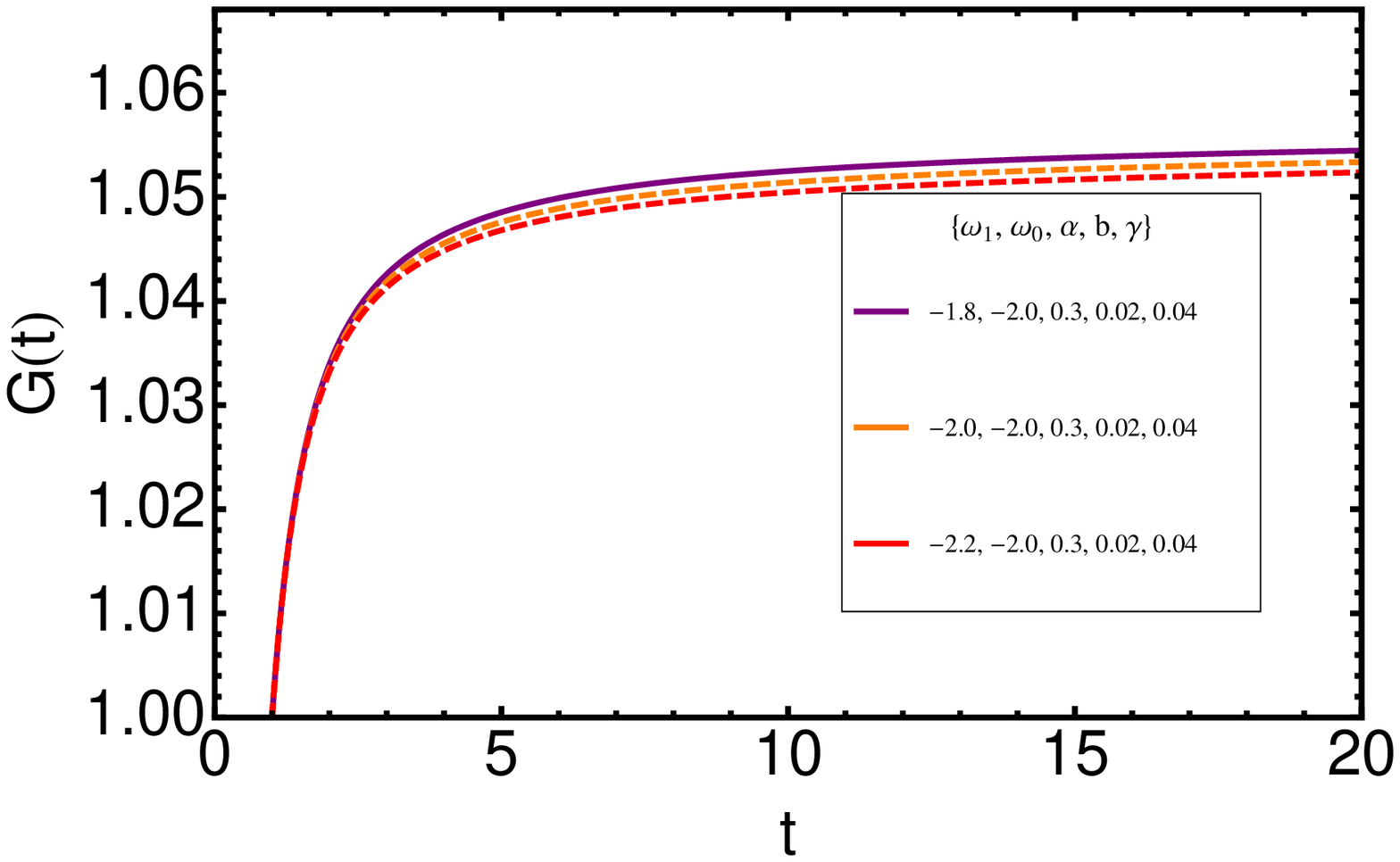} &
\includegraphics[width=50 mm]{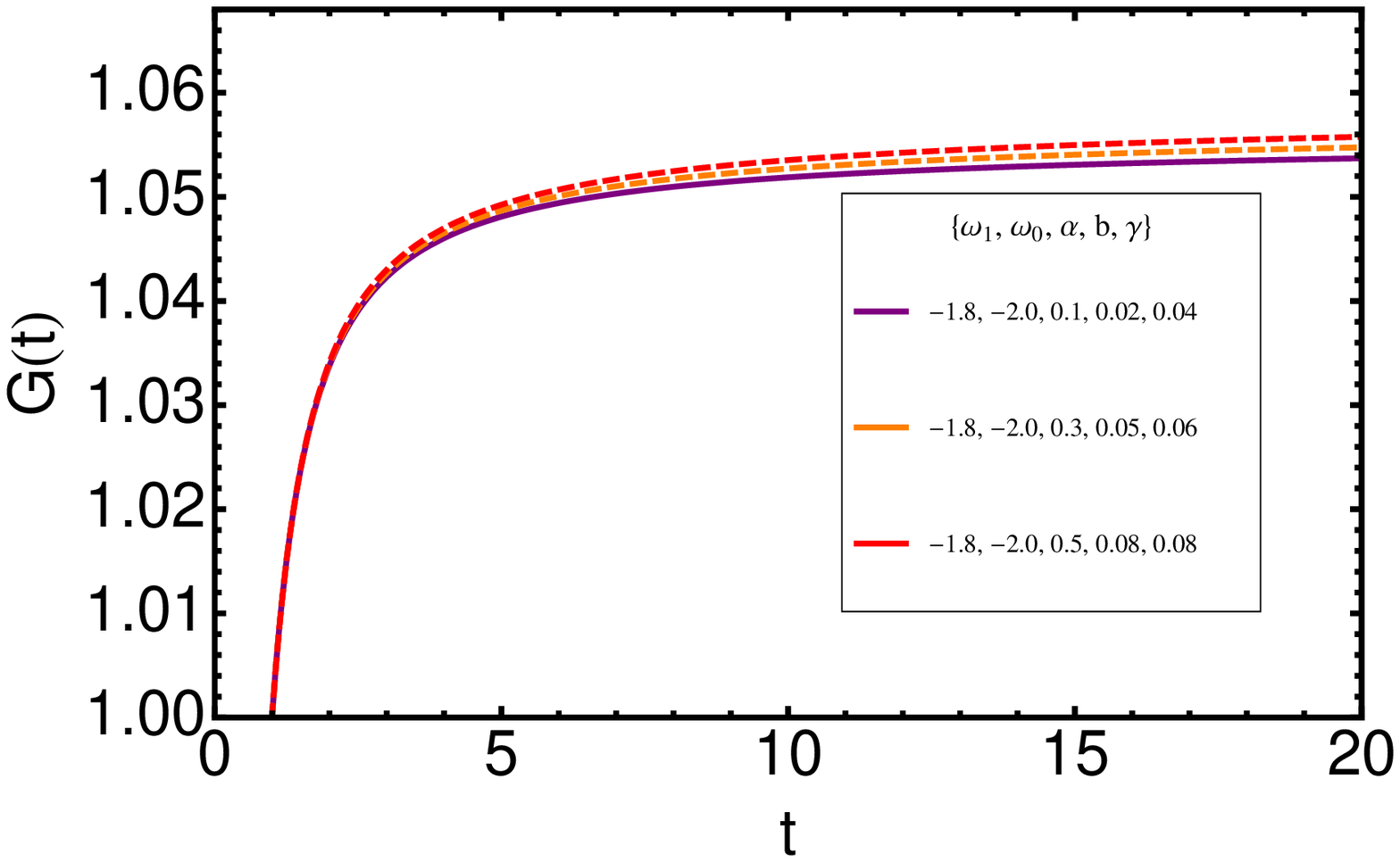}
 \end{array}$
 \end{center}
\caption{Interacting two components}
 \label{fig:4}
\end{figure}

\begin{figure}[h]
 \begin{center}$
 \begin{array}{cccc}
 \includegraphics[width=60 mm]{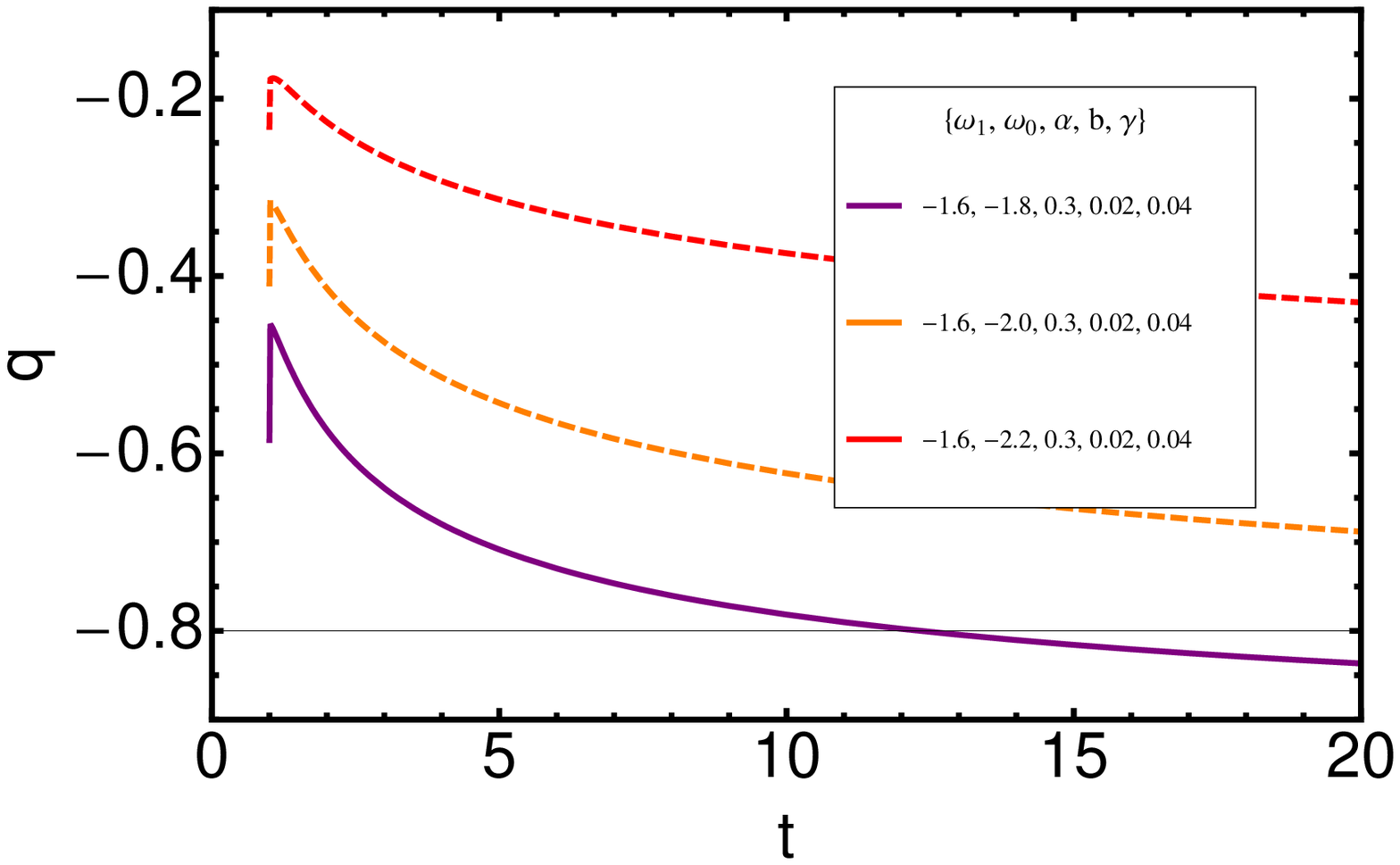} &
 \includegraphics[width=60 mm]{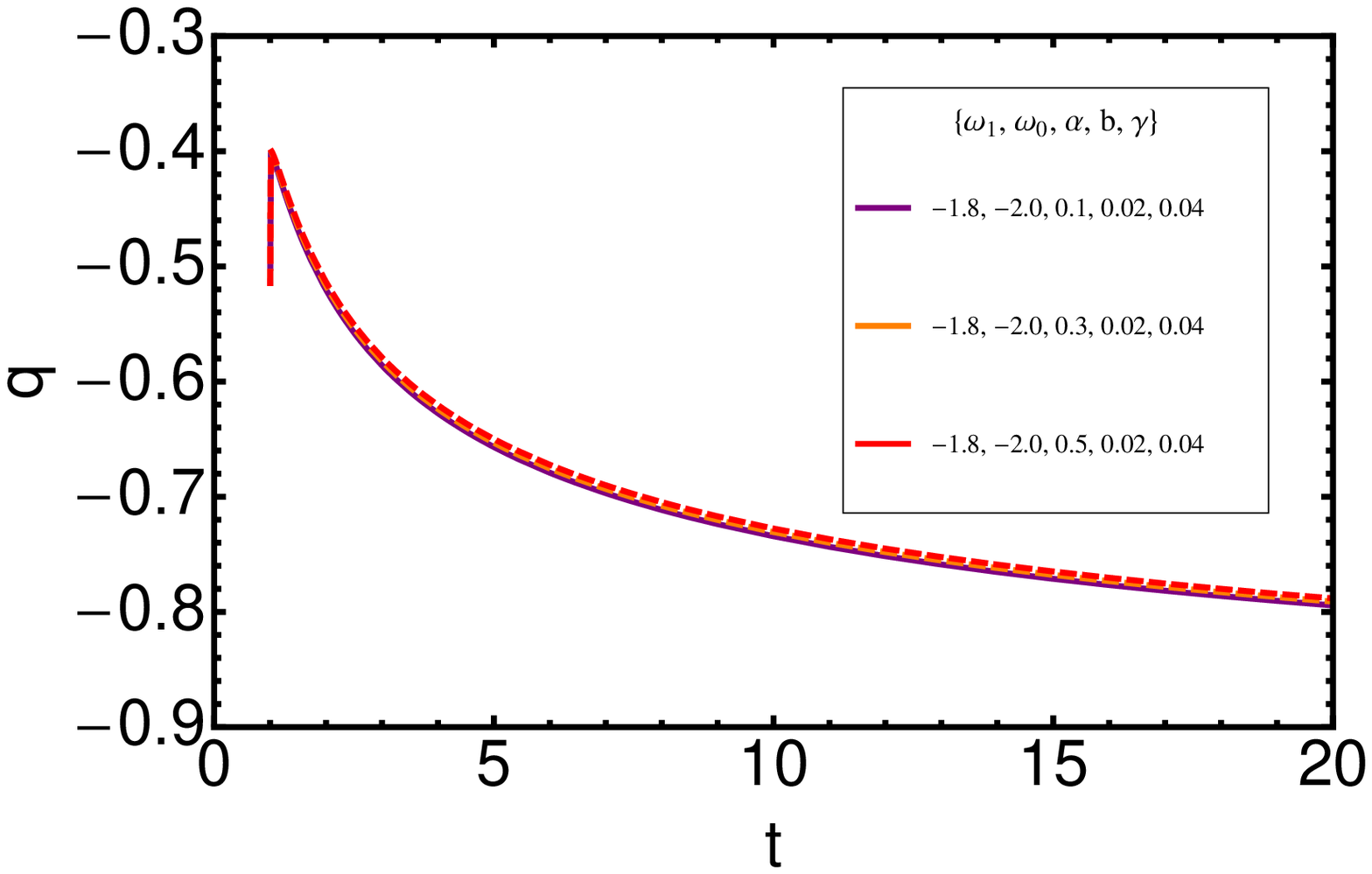} \\
 \includegraphics[width=60 mm]{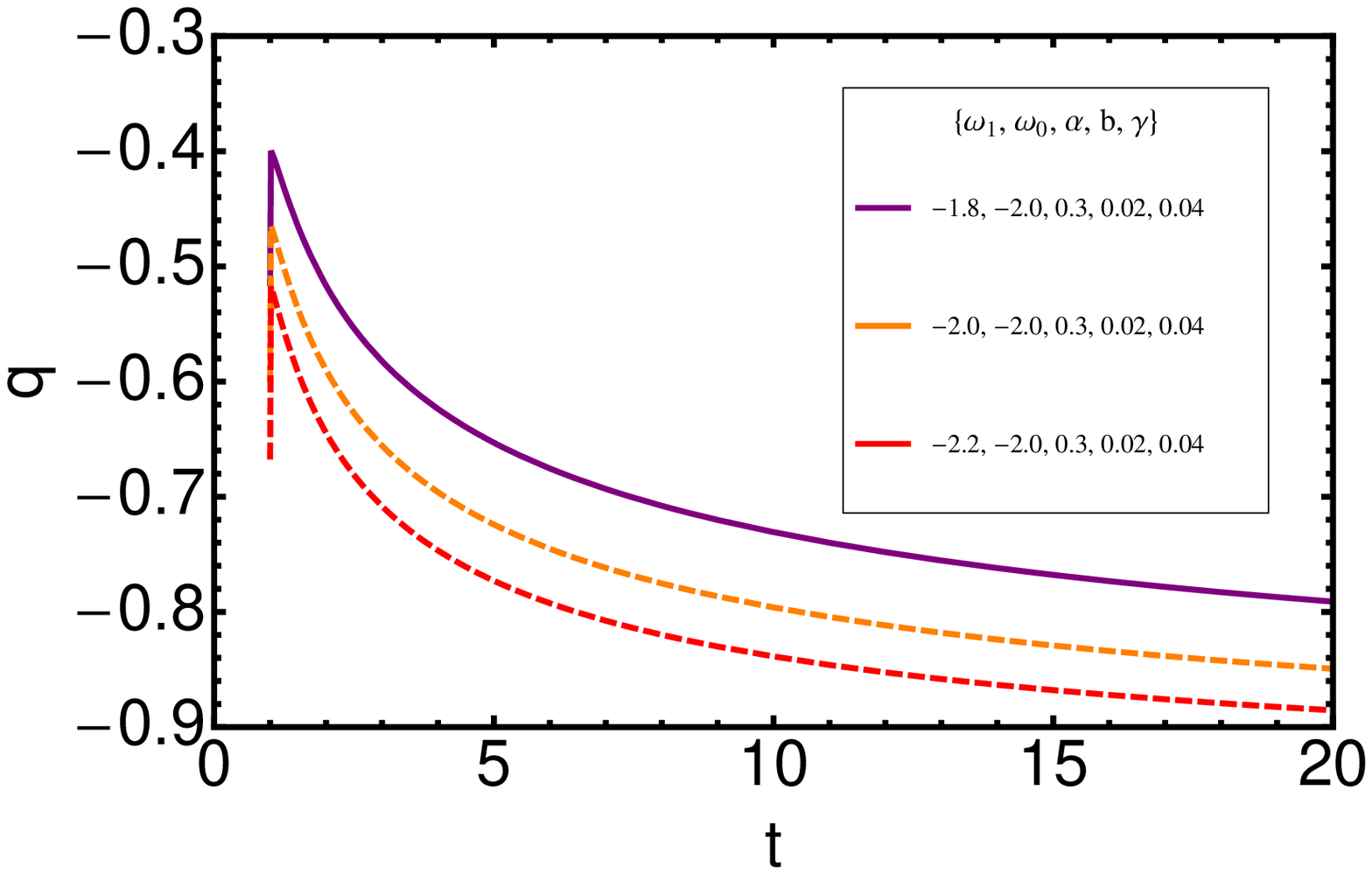} &
 \includegraphics[width=60 mm]{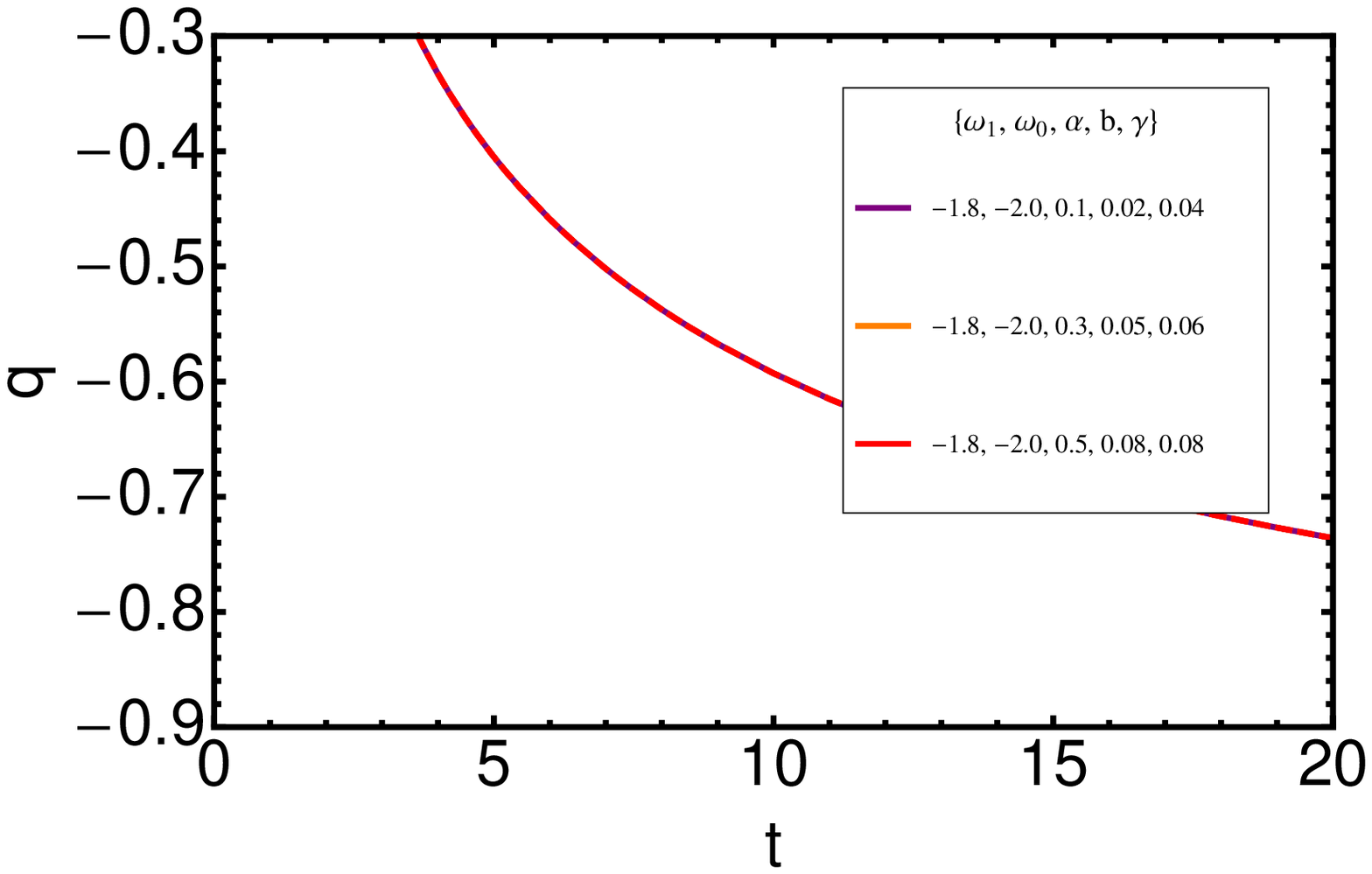}
 \end{array}$
 \end{center}
\caption{interacting two components}
 \label{fig:5}
\end{figure}

\begin{figure}[h]
 \begin{center}$
 \begin{array}{cccc}
 \includegraphics[width=60 mm]{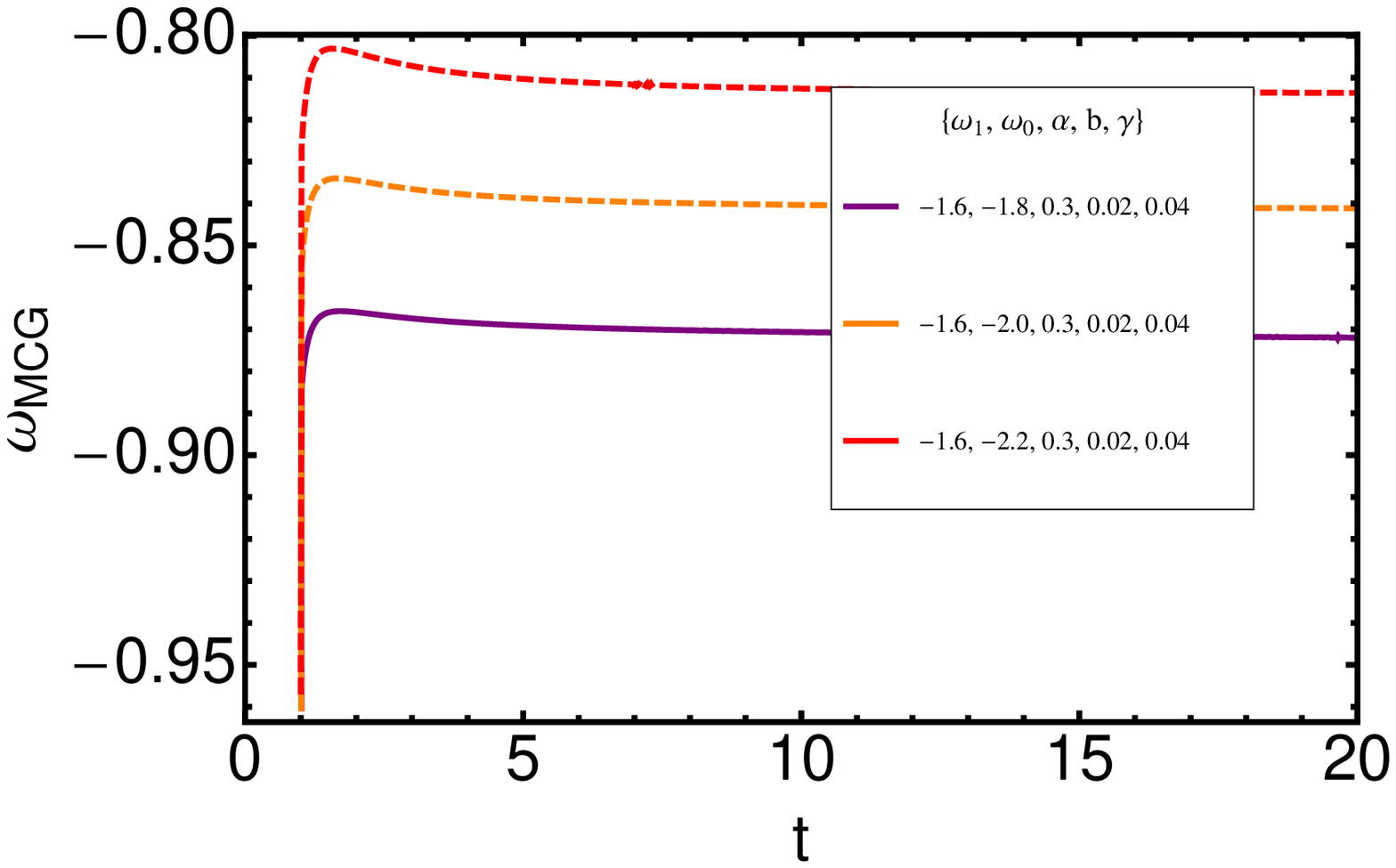} &
 \includegraphics[width=60 mm]{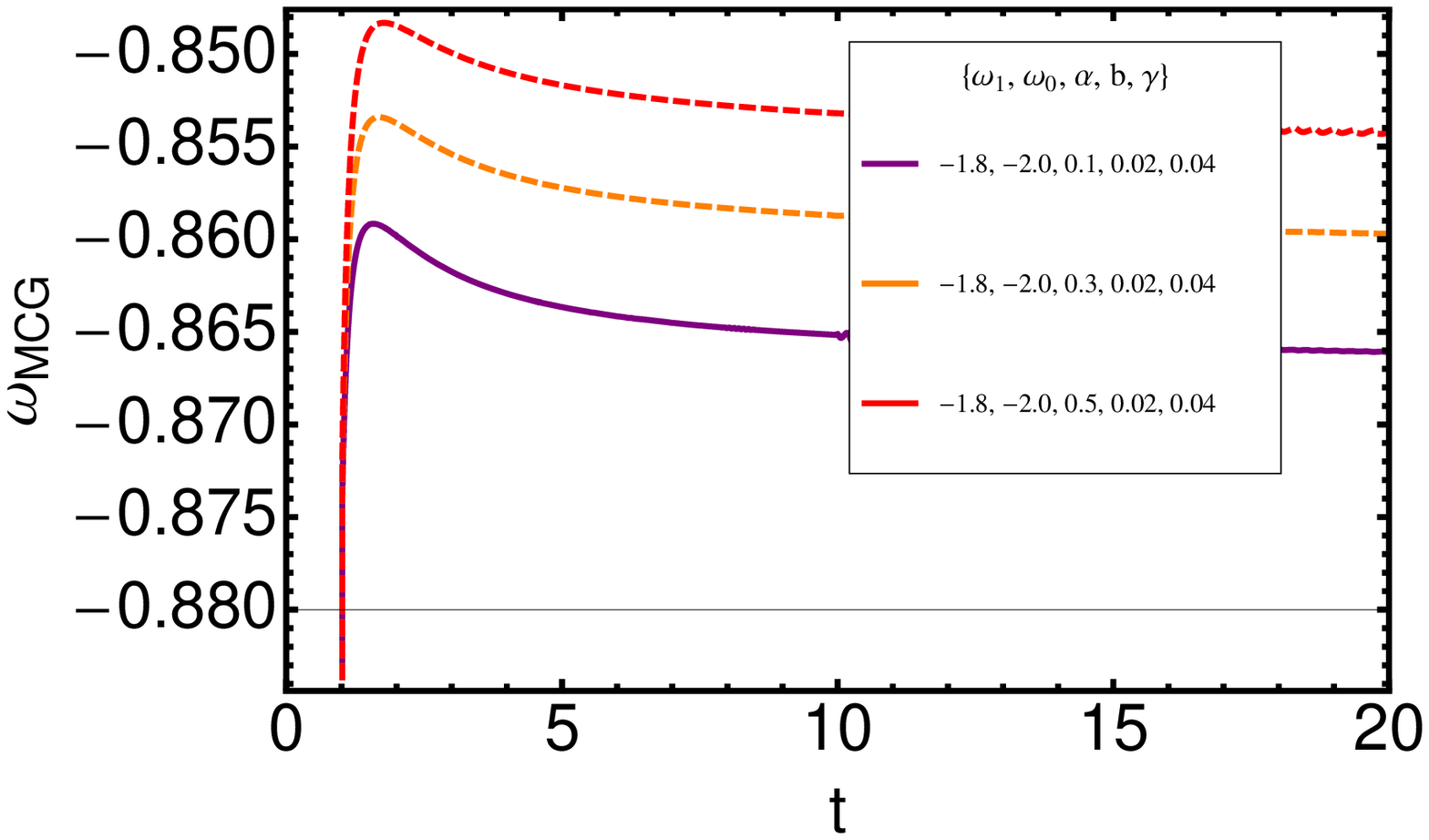} \\
 \includegraphics[width=60 mm]{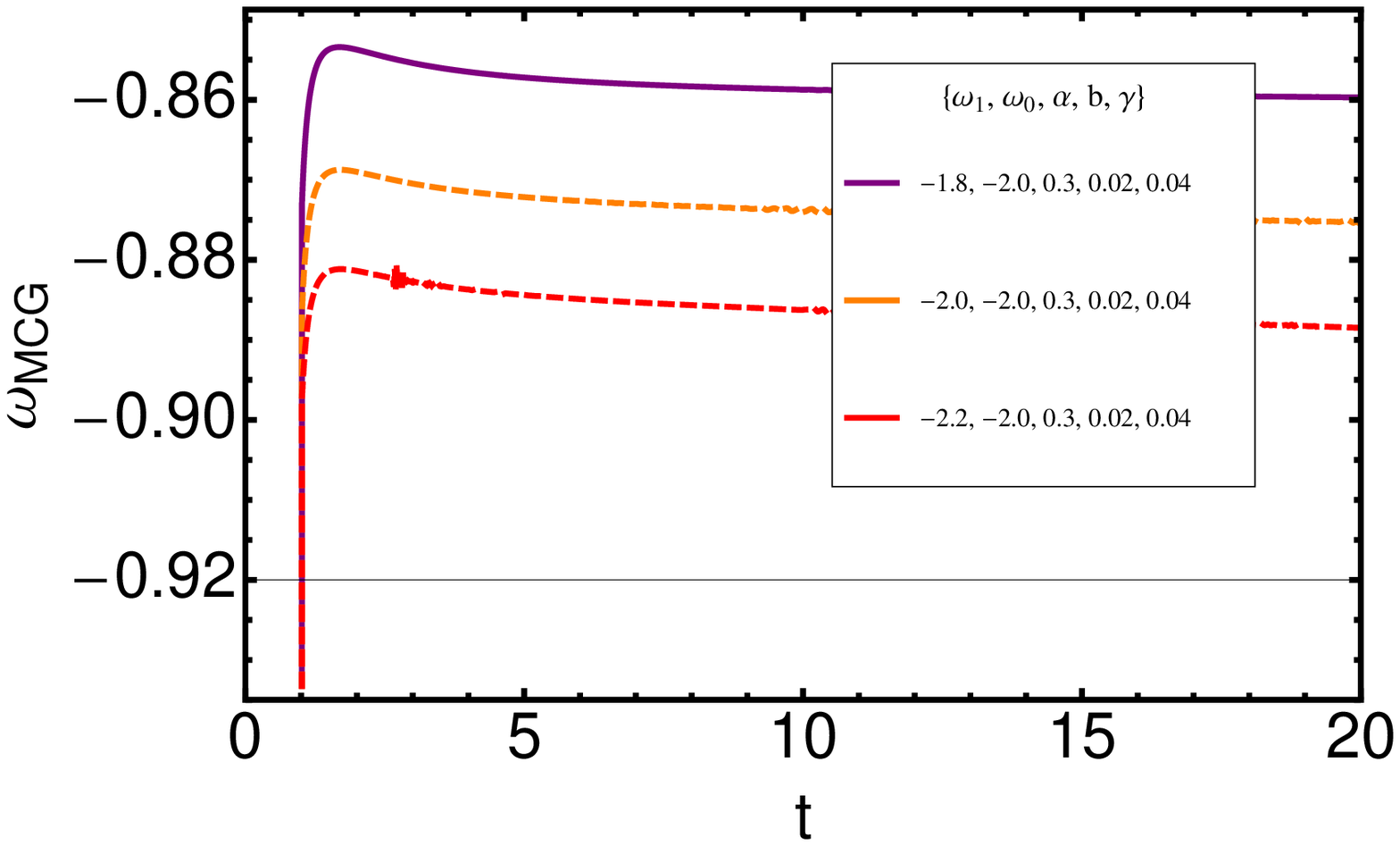} &
 \includegraphics[width=60 mm]{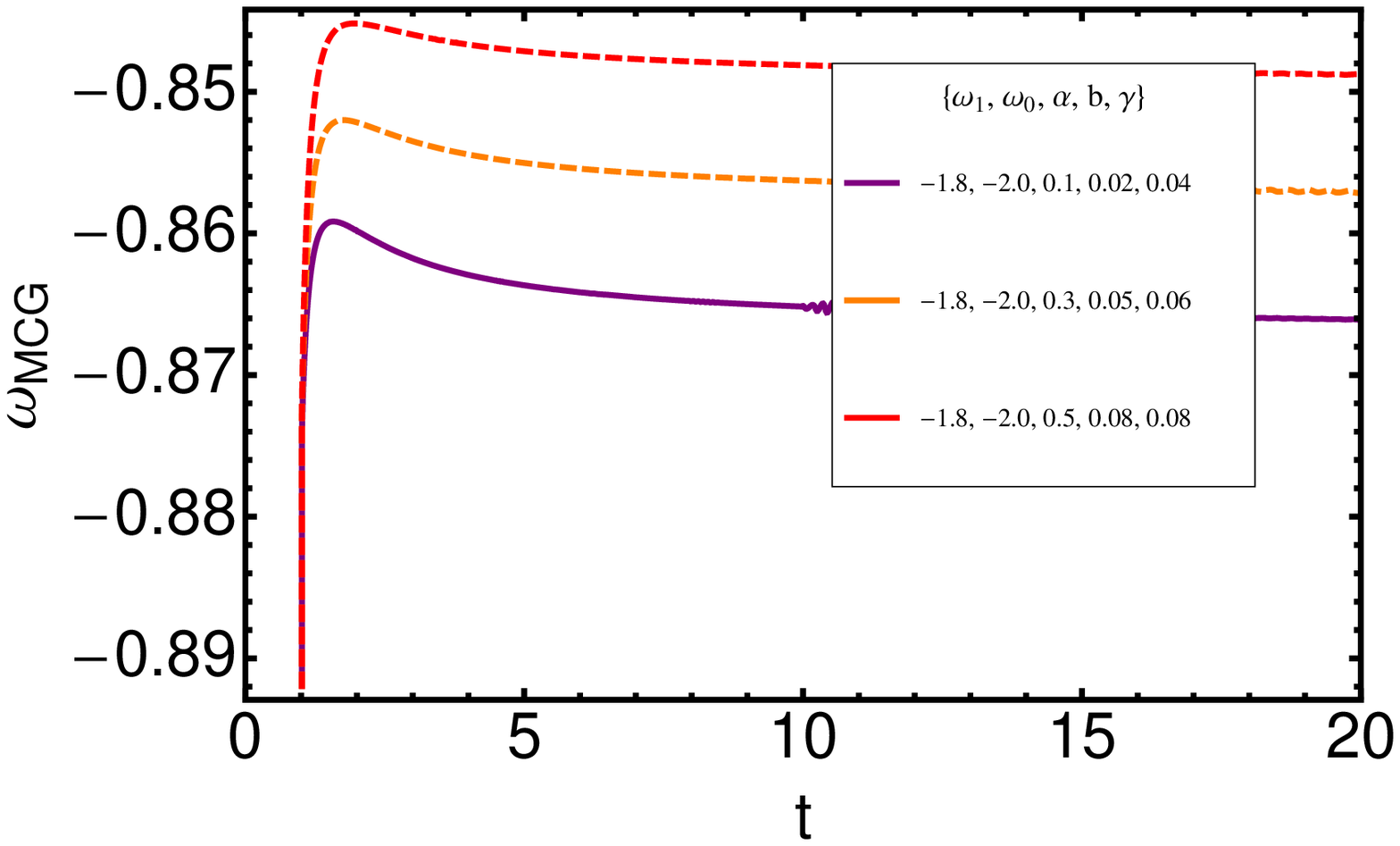}
 \end{array}$
 \end{center}
 \caption{Interacting two components}
 \label{fig:6}
\end{figure}

\section*{\large{Discussion}}
In this paper we consider Modified Chaplygin gas as a cosmological
model which unifies dark matter and dark energy. This work which is
extended version of previous works [1-3] written based on
interesting idea which tells that unknown physics separates dark
side of Universe to gas and fluid which may be interact with each
other while there is not any interaction between remaining darkness
of Universe and fluid. Such interaction yields to varying modified
Chaplygin gas. This assumption allows us still to think that our
Universe consists of mixture of a fluid and a varying modified
Chaplygin gas. An important point of this paper is
that we considered variable $G$ and $\Lambda$.\\
First of all we studied Universe with variable $G$ and $\Lambda$ and
varying modified Chaplygin gas as a simple model. Then we consider
two component fluid which involve sign-changeable interaction. We
assumed $\Lambda(t)$ proportional to $t^{-2}$ and have numerical
analysis of $G(t)$, declaration parameter and $\omega_{MCG}$. In the
first case we found that the variation of $\alpha$ is not important
for $G(t)$. However we found that $G(t)$ is increasing function of
time. The declaration parameter increased at the early stage ant
then is decreasing function of time. $\omega_{MCG}$ has similar
behavior as $q$. In the second case where we consider
sign-changeable interaction $G(t)$ is also increasing function of
time but yields to a constant at the late time. The declaration
parameter of this case has similar behavior of the previous case. It
is seen that the Universe initially undergoes a rapidly falling
acceleration followed by a rise in it. At a particular epoch the
Universe get into a phase of constant acceleration in which we are
presently located. Finally $\omega_{MCG}$ increased at the initial
stage and rich to a maximum and finally yields to a constant at the late time.\\
We can also investigate other cosmological quantities such as scale
factor, $\omega(t)$ and density. In the Figs. 7, 8 and 9 we draw
these parameters for the case of single component fluid. The first
plot of the Fig. 7 shows that increasing $|\omega_{0}|$ decreased
scale factor to a constant. As we can see from the last plot this
constant obtained for $\omega_{0}=-2.2$, $\omega_{1}=-1.8$ and
$\alpha=0.3$. The second plot of the Fig. 7 shows that the scale
factor decreased by increasing of $\alpha$. on the other hand plots
of the Fig. 9 show that energy density is decreasing function of
time.\\
Scale factor of two components fluid Universe drawn in the plots of
the Fig. 10. It shows that increasing $\omega_{0}$ decreased scale
factor. The last graph of the Fig. 10 is represent general case for
the behavior of $a$. Plots of the Fig. 11 show that EoS parameter
yields to a negative constant at the late time. Finally plots of the
Fig. 12 show that energy density is decreasing function of time. In
that case the second plot tells that increasing $\alpha$ decreased
the energy density.\\
In this paper we considered modified Chaplygin gas which may be
extended to the case of modified cosmic Chaplygin gas with the
following equation of state [70],
\begin{equation}\label{s24}
p=\mu\rho-\frac{1}{\rho^{\alpha}}\left[\frac{B}{1+\omega}-1+(\rho^{1+\alpha}-\frac{B}{1+\omega}+1)^{-\omega}\right],
\end{equation}
where $\omega$ is the cosmic parameter. Also one can include shear
viscosity and bulk viscosity [76] and investigate effect of them on
cosmological parameters.
\section*{Acknowledgments}
Martiros Khurshudyan has been supported by EU fonds in the frame of the program FP7-Marie Curie Initial Training Network INDEX NO.289968.

\newpage
\section*{Appendix: Single Component fluid Universe}
\begin{figure}[h]
 \begin{center}$
 \begin{array}{cccc}
\includegraphics[width=50 mm]{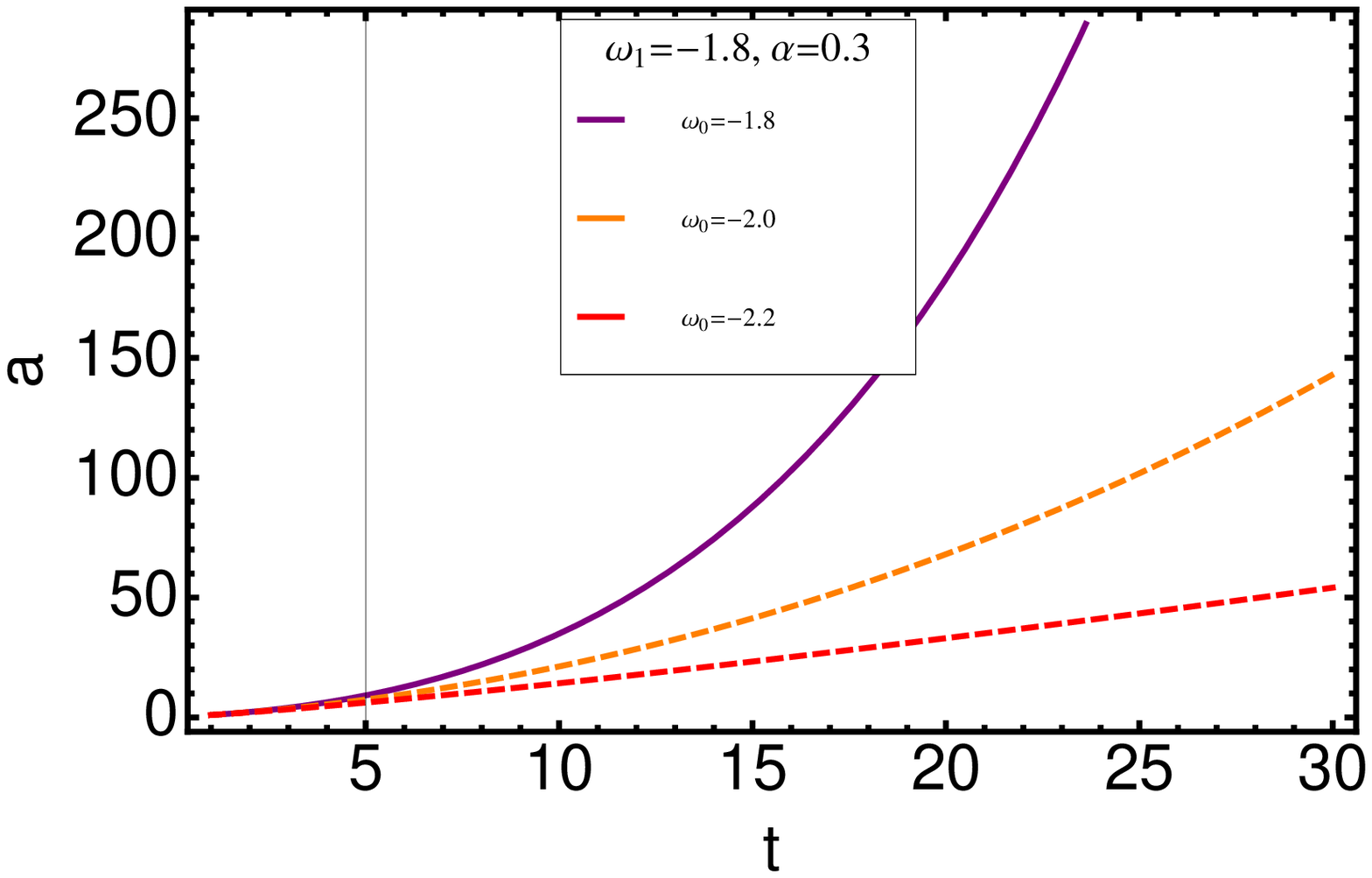} &
\includegraphics[width=50 mm]{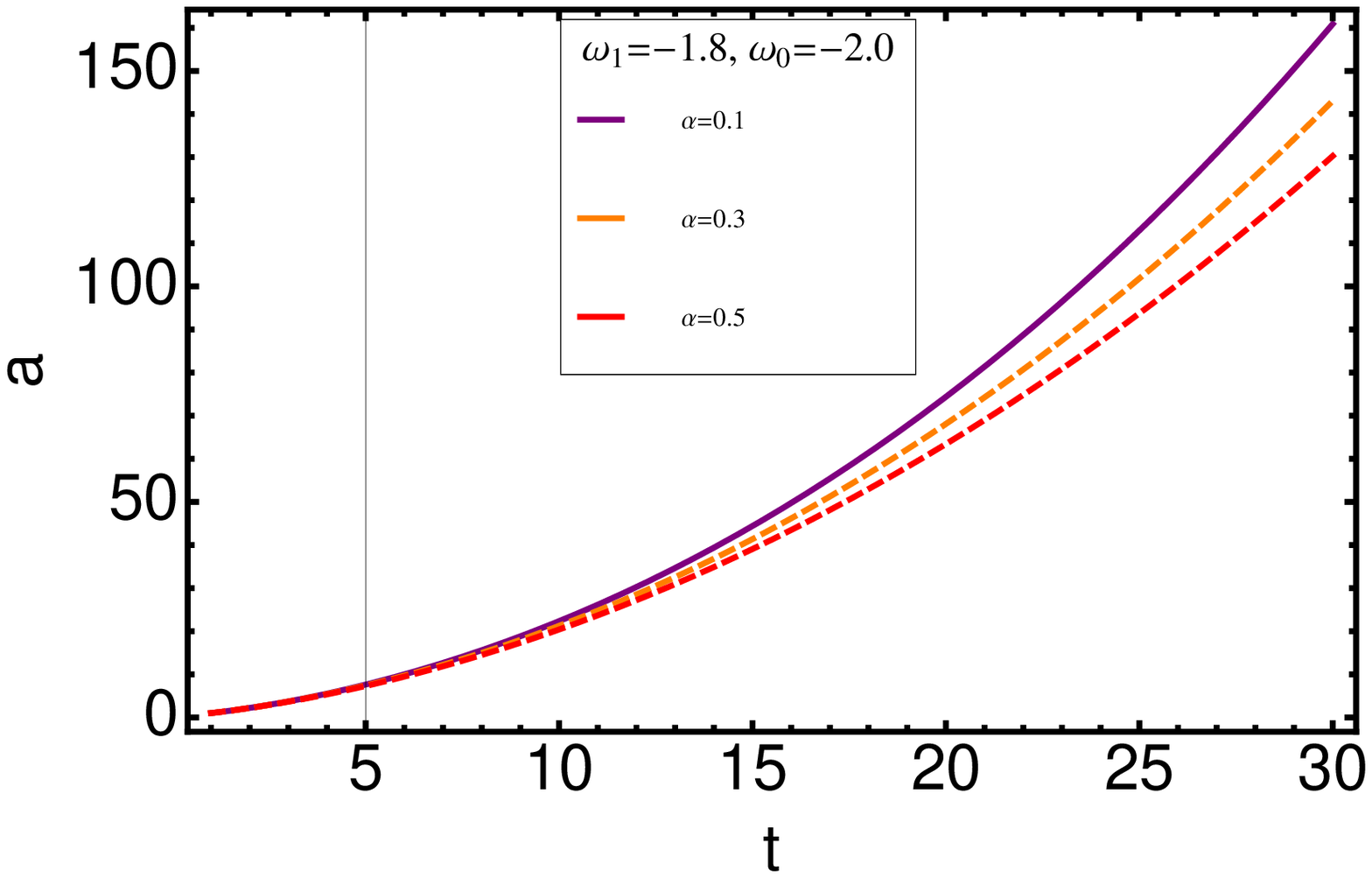}\\
\includegraphics[width=50 mm]{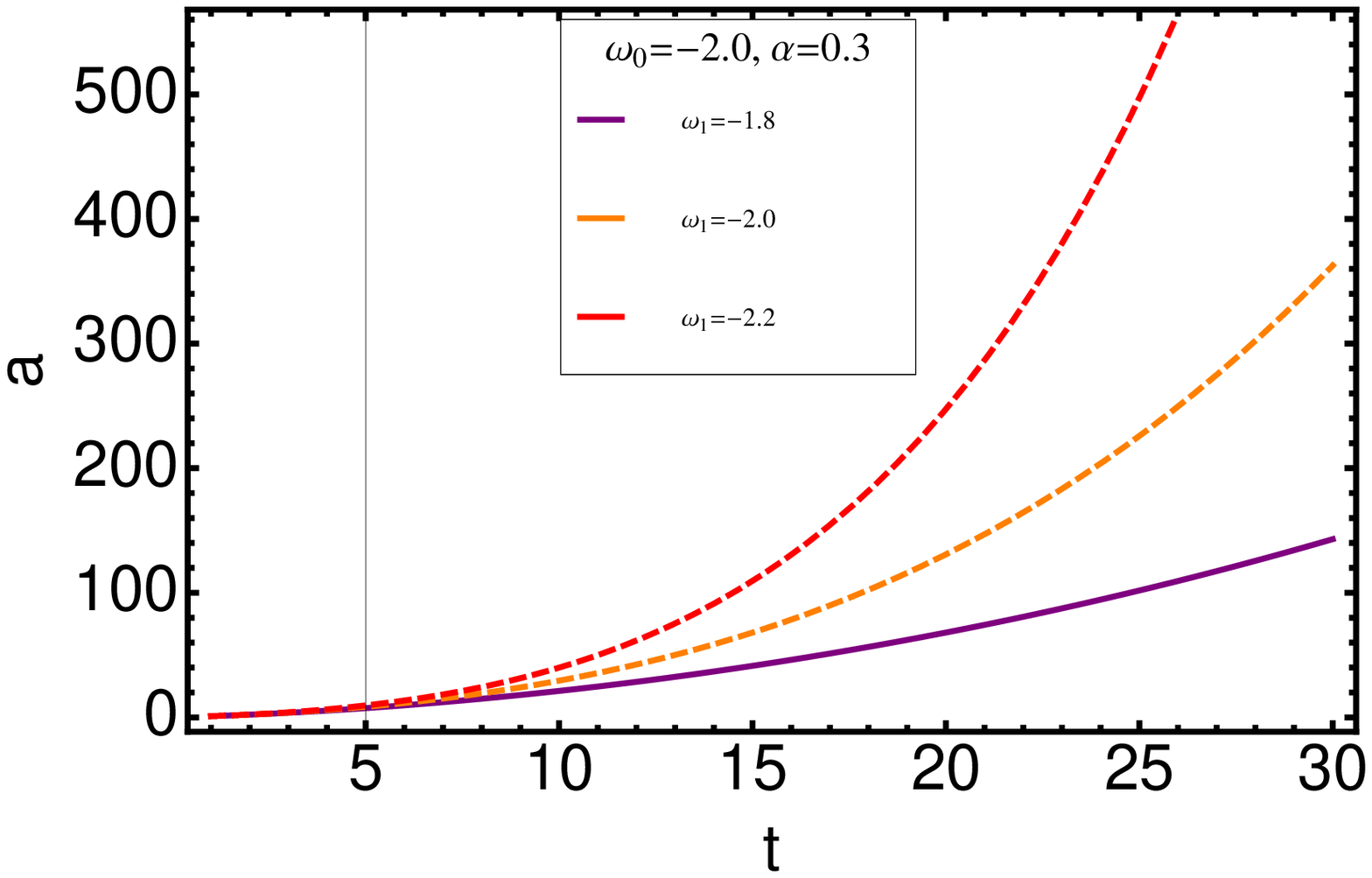}&
\includegraphics[width=50 mm]{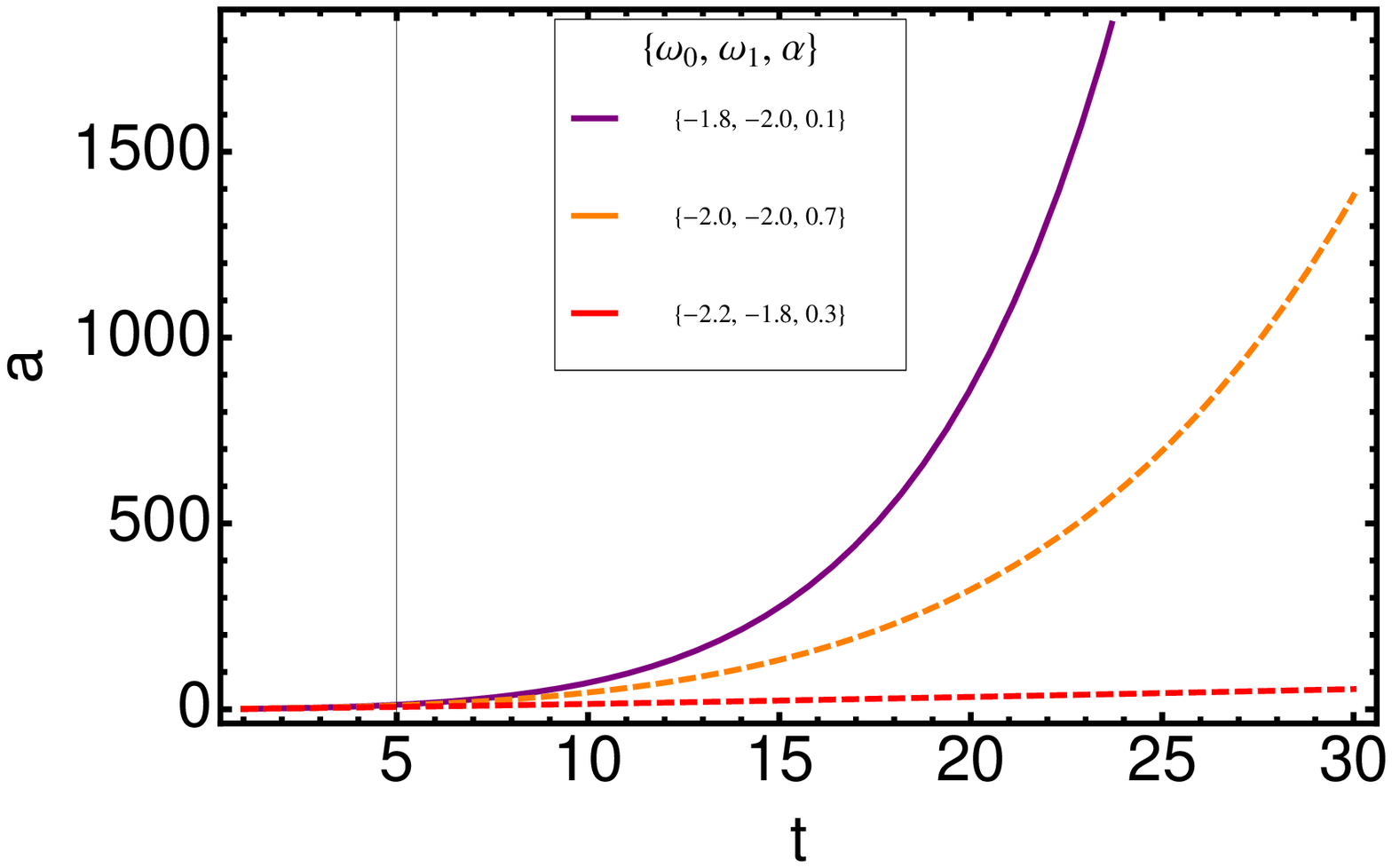}
 \end{array}$
 \end{center}
\caption{Scale factor of single component fluid Universe}
 \label{fig:7}
\end{figure}

\begin{figure}[h]
 \begin{center}$
 \begin{array}{cccc}
\includegraphics[width=50 mm]{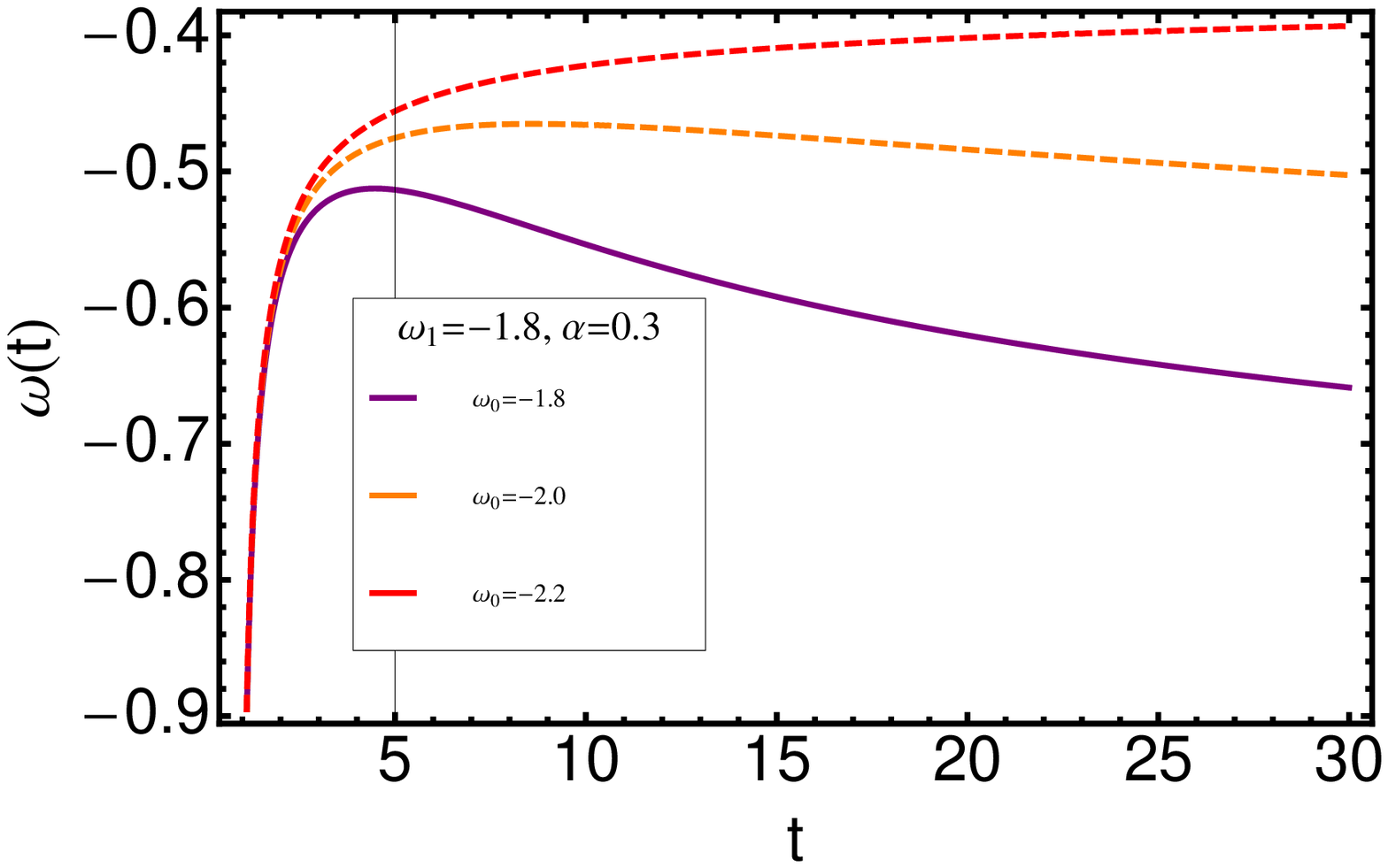} &
\includegraphics[width=50 mm]{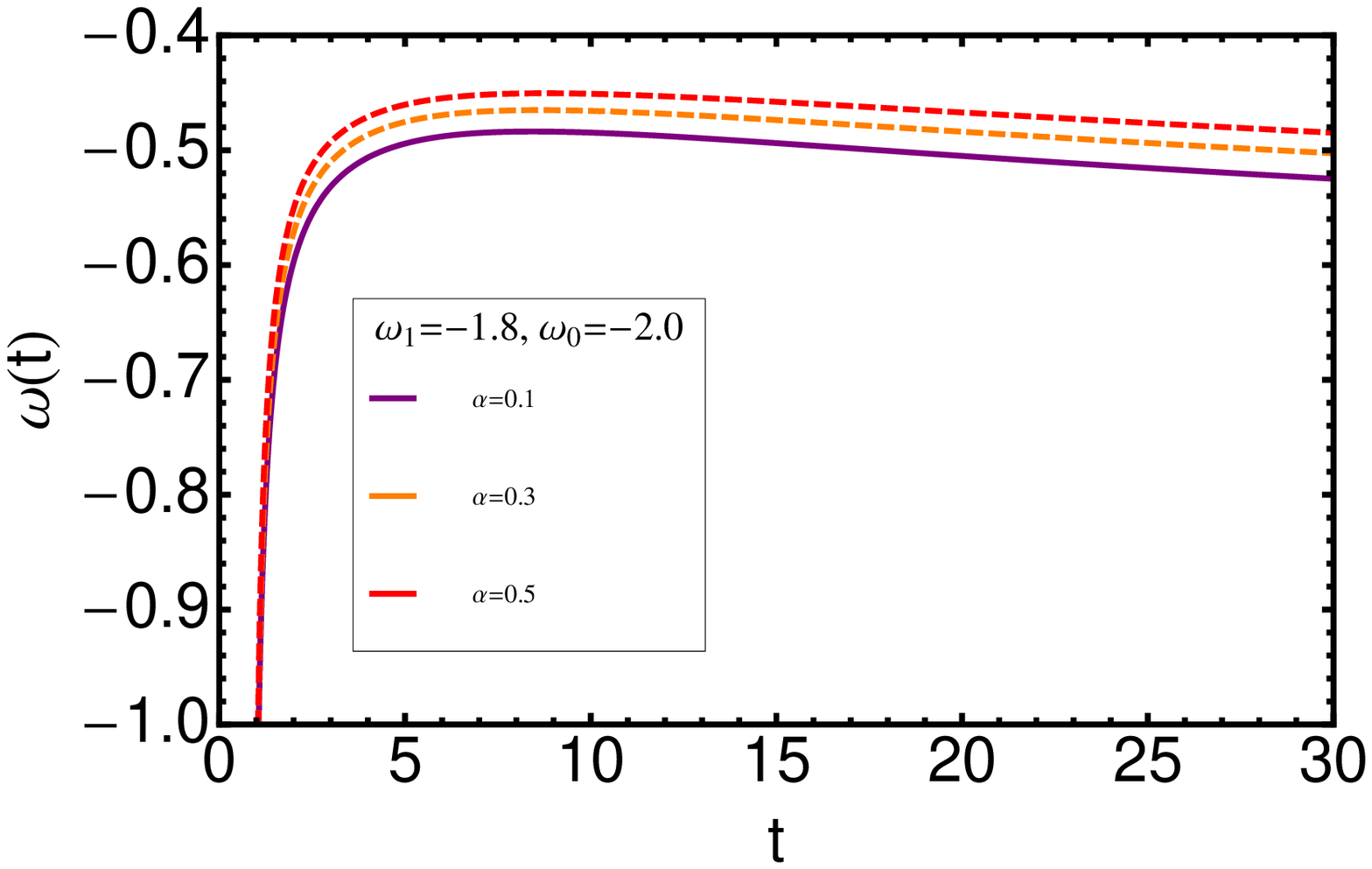}\\
\includegraphics[width=50 mm]{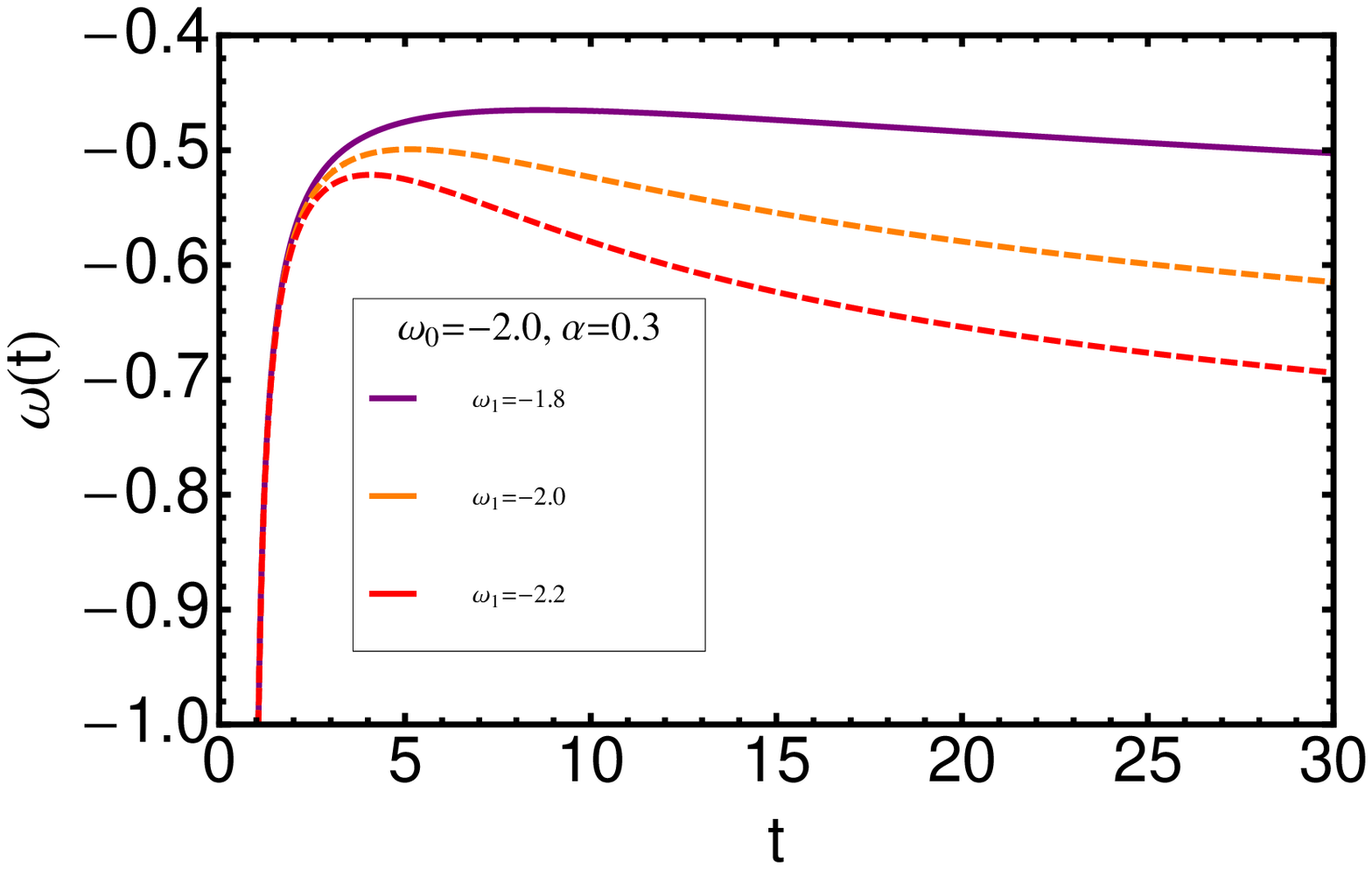} &
\includegraphics[width=50 mm]{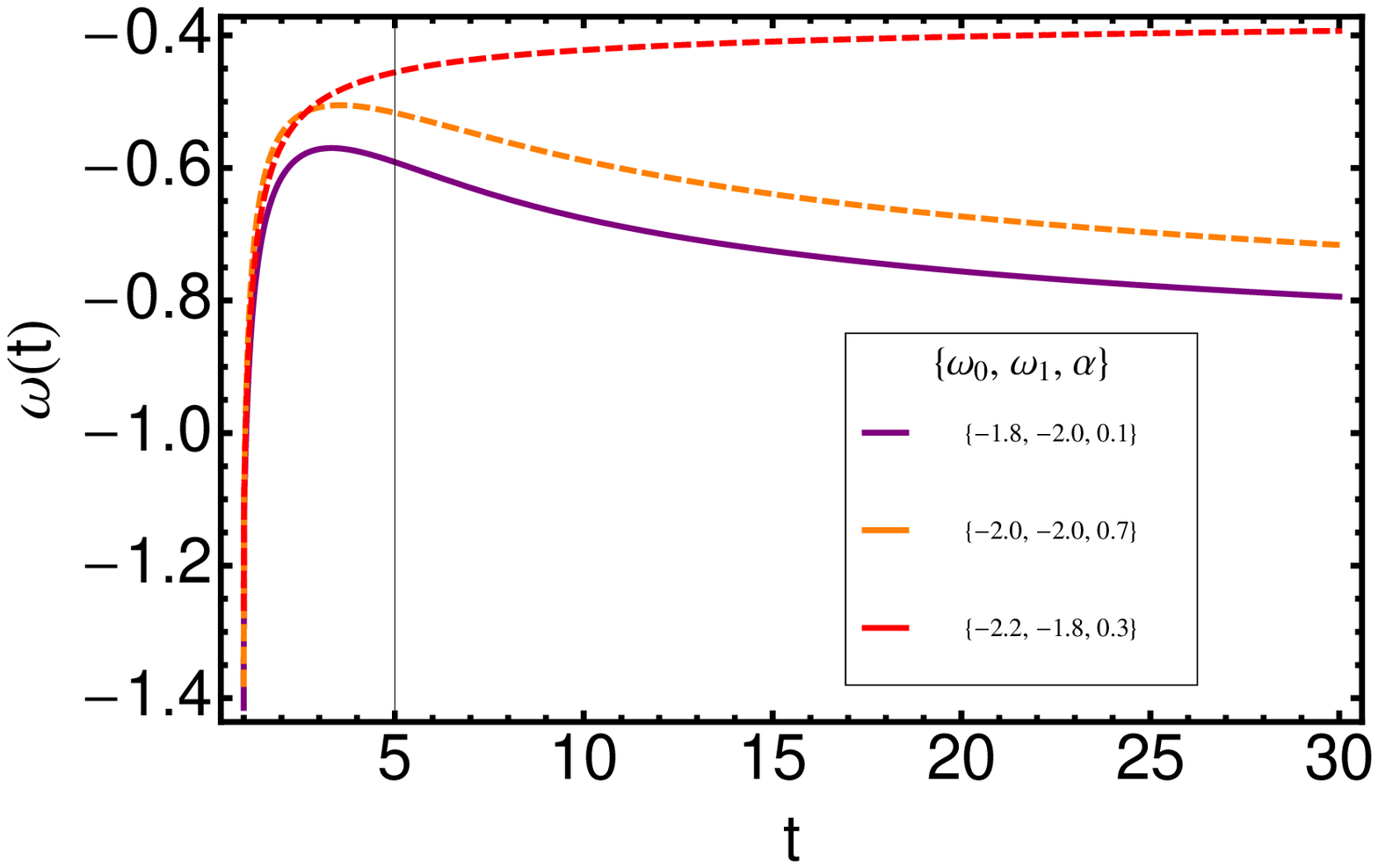}
 \end{array}$
 \end{center}
\caption{EoS parameter of single component fluid Universe}
 \label{fig:8}
\end{figure}

\begin{figure}[h]
 \begin{center}$
 \begin{array}{cccc}
 \includegraphics[width=60 mm]{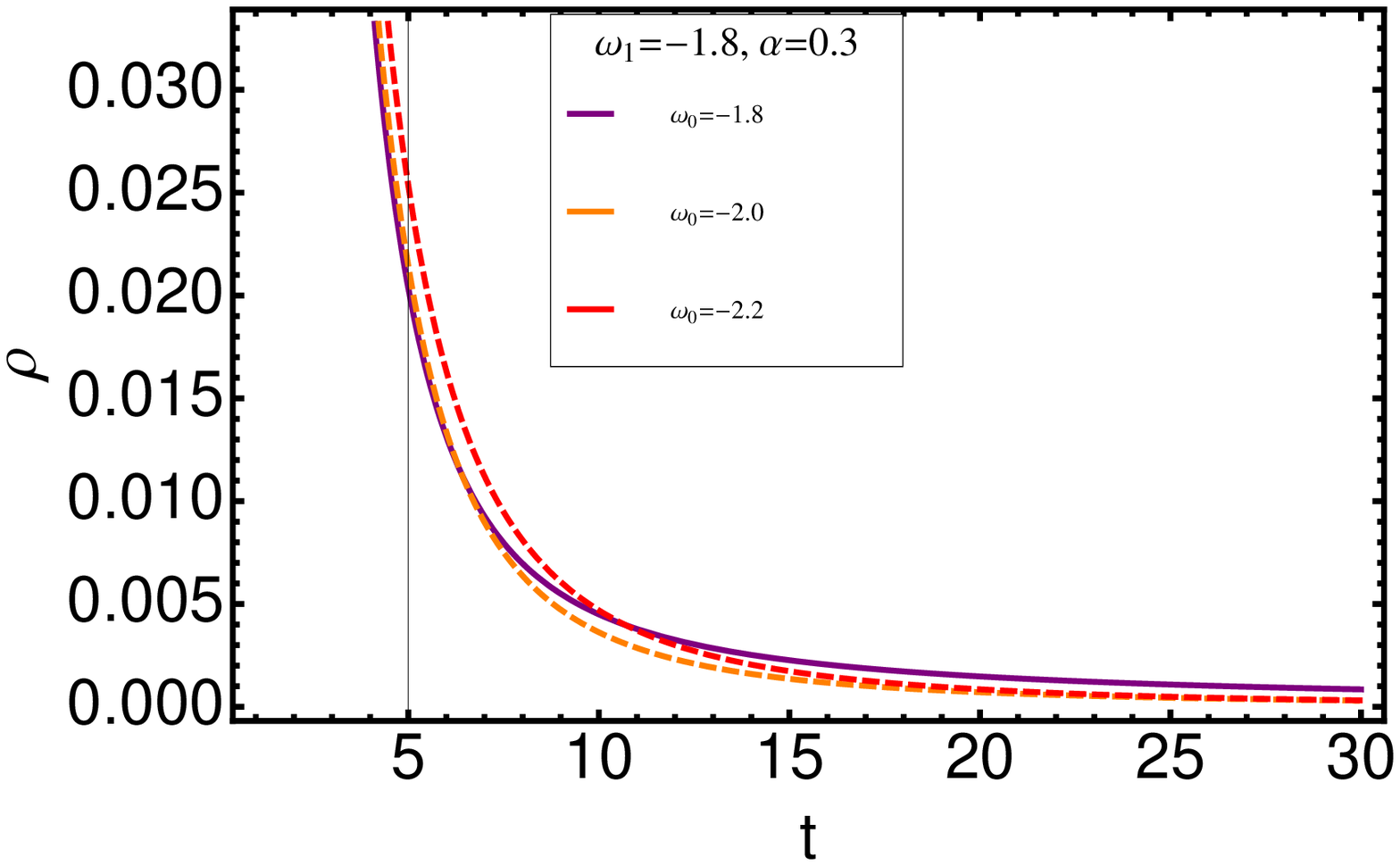} &
 \includegraphics[width=60 mm]{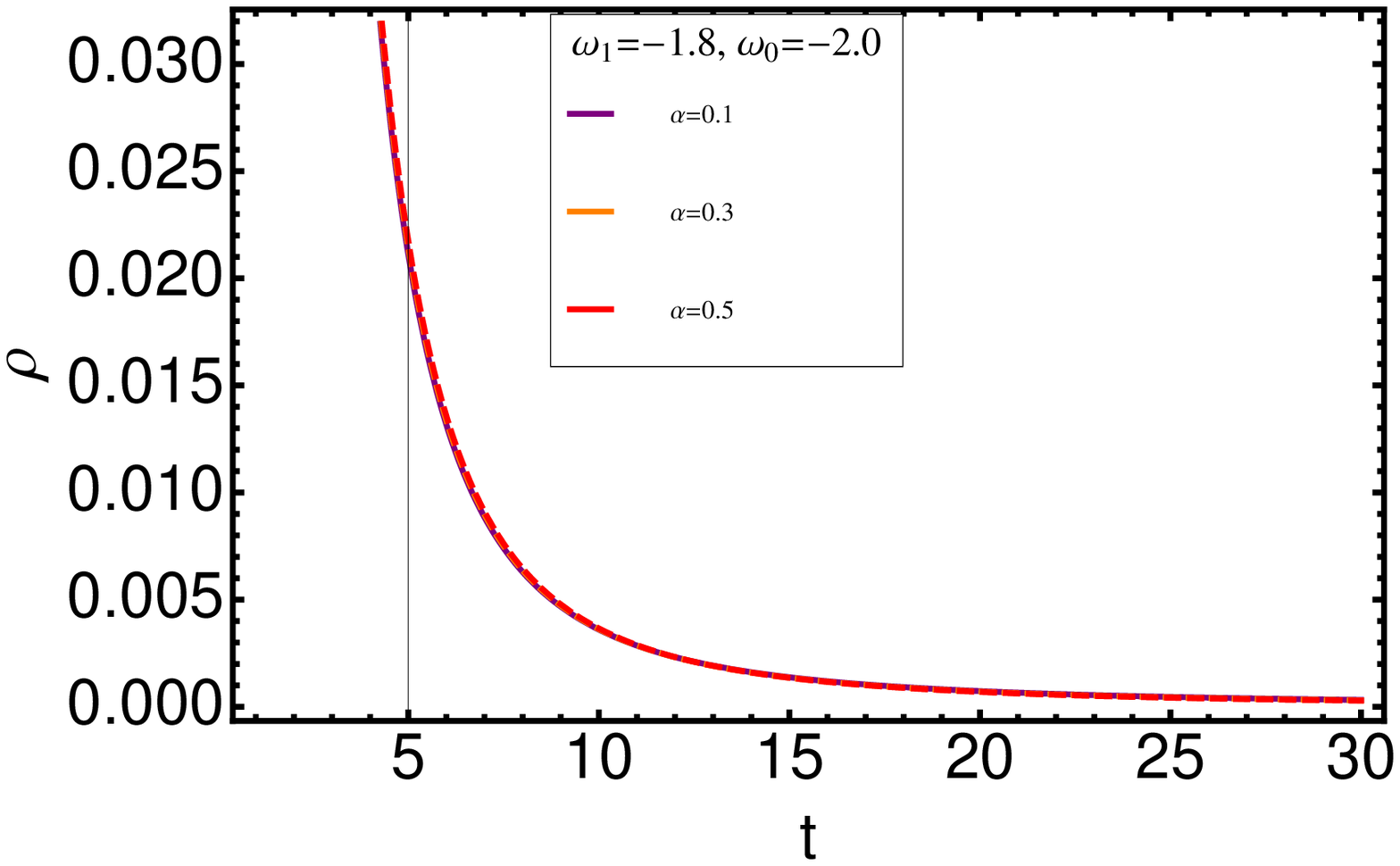} \\
 \includegraphics[width=60 mm]{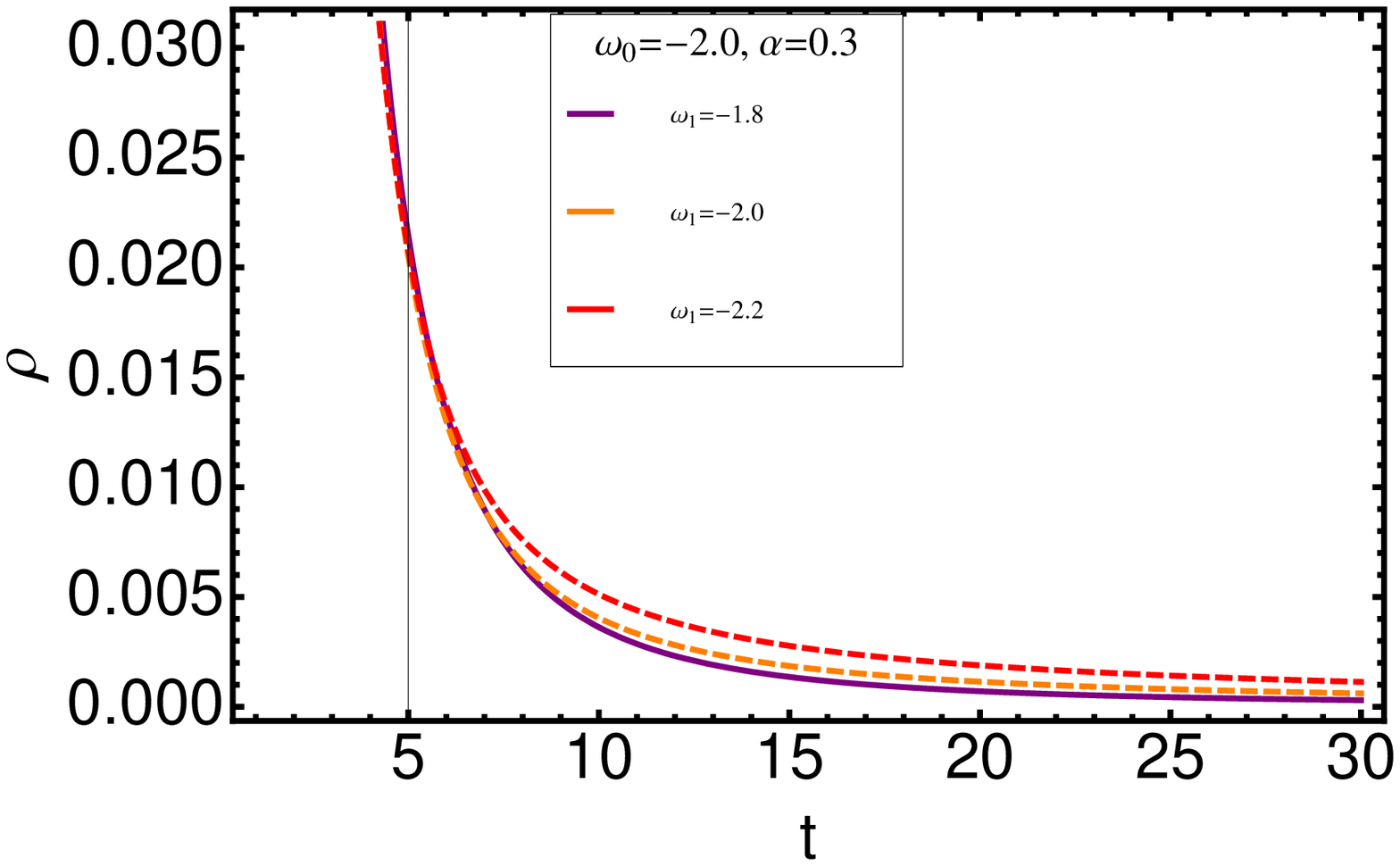} &
 \includegraphics[width=60 mm]{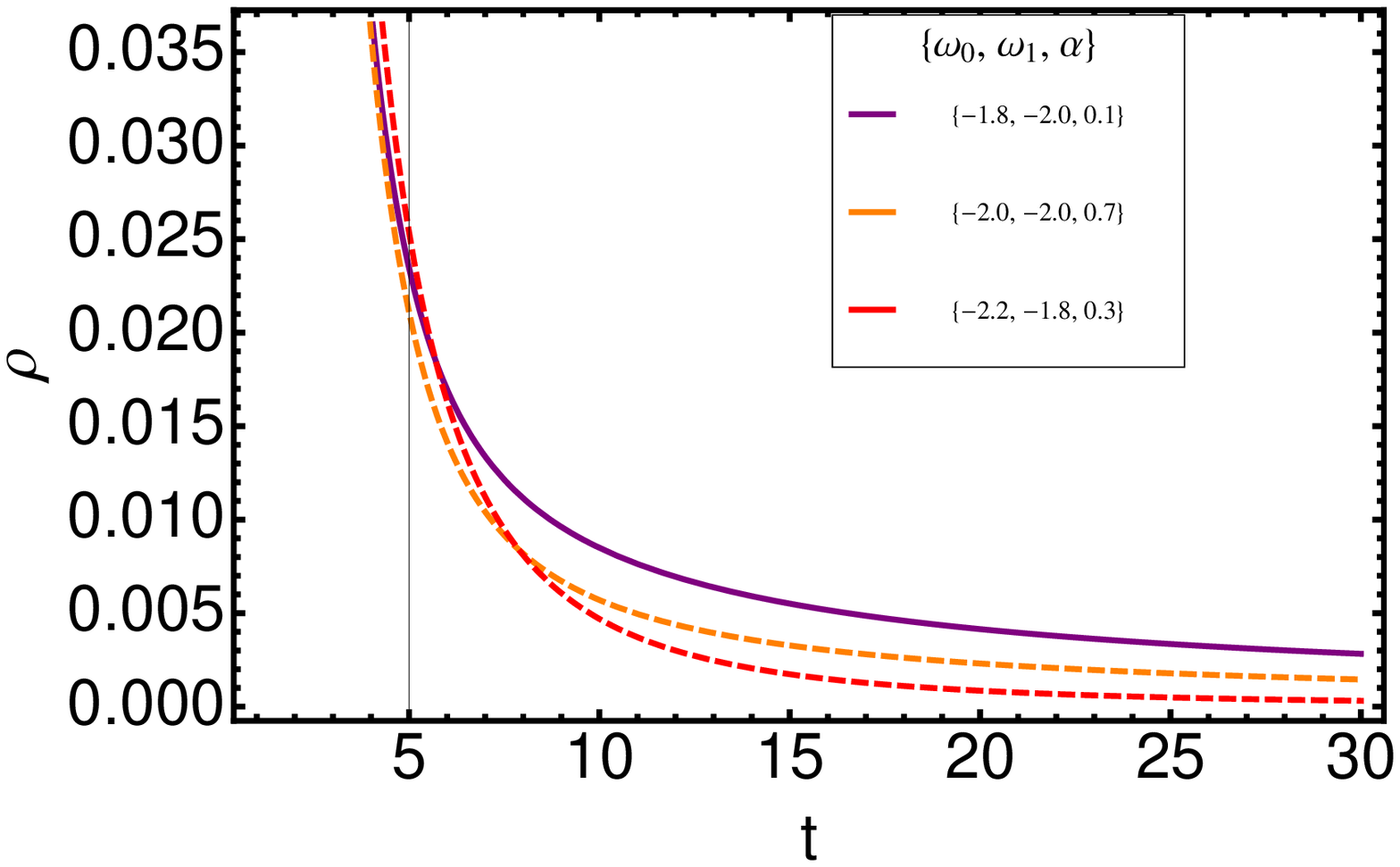}
 \end{array}$
 \end{center}
 \caption{Energy density of single component fluid Universe}
 \label{fig:9}
\end{figure}
\newpage
\section*{Appendix: Two Components fluid Universe}
\begin{figure}[h]
 \begin{center}$
 \begin{array}{cccc}
\includegraphics[width=50 mm]{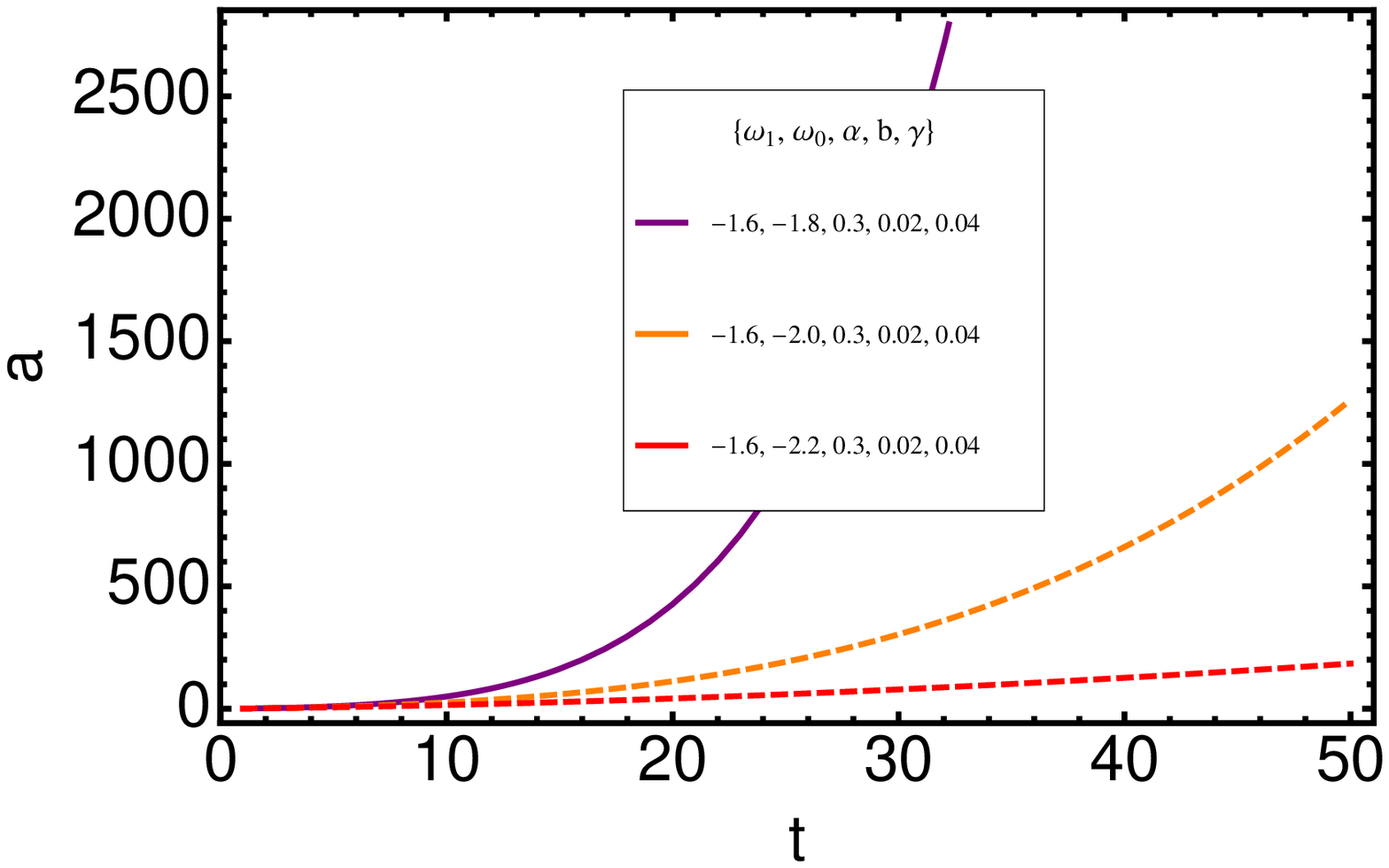} &
\includegraphics[width=50 mm]{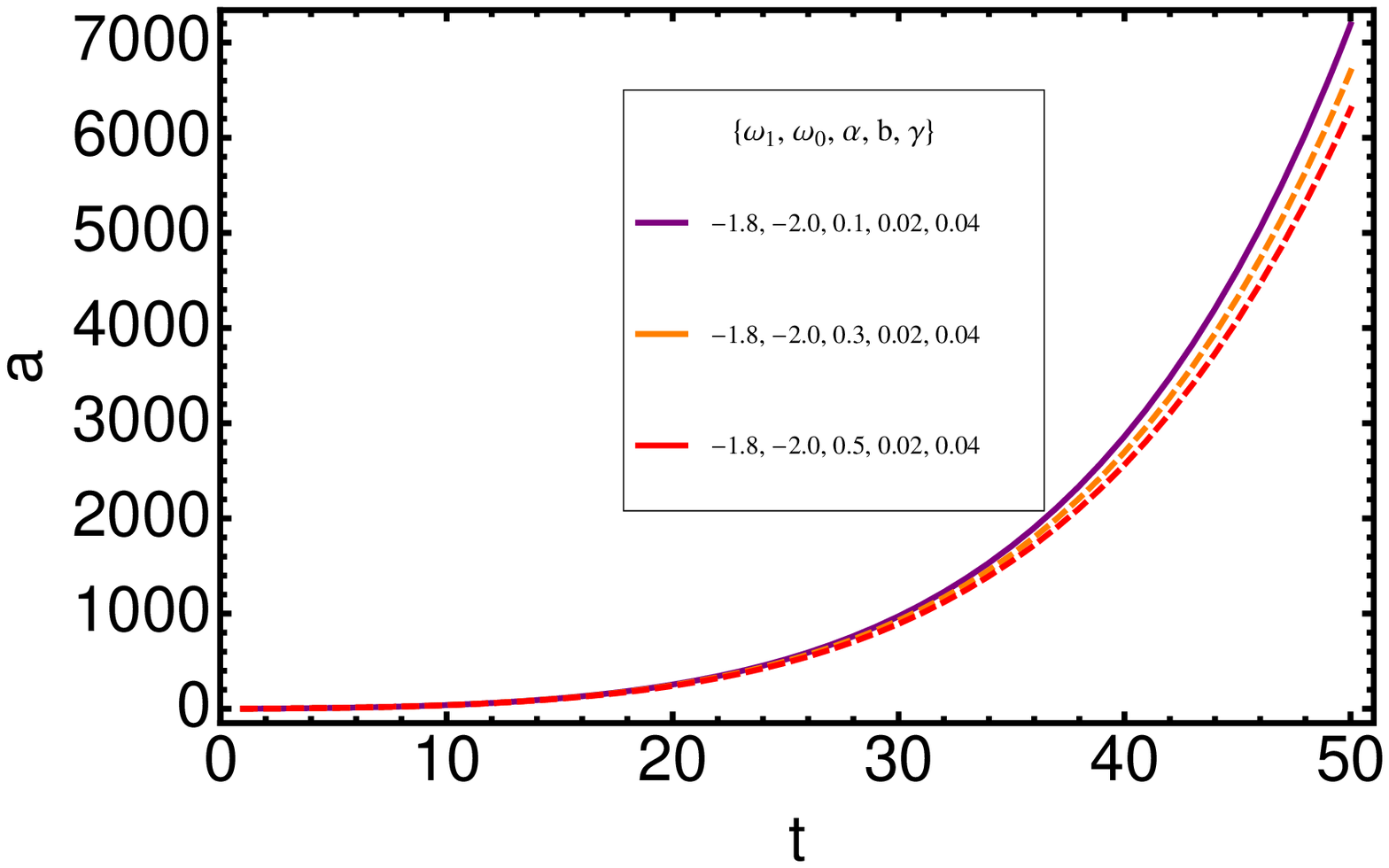}\\
\includegraphics[width=50 mm]{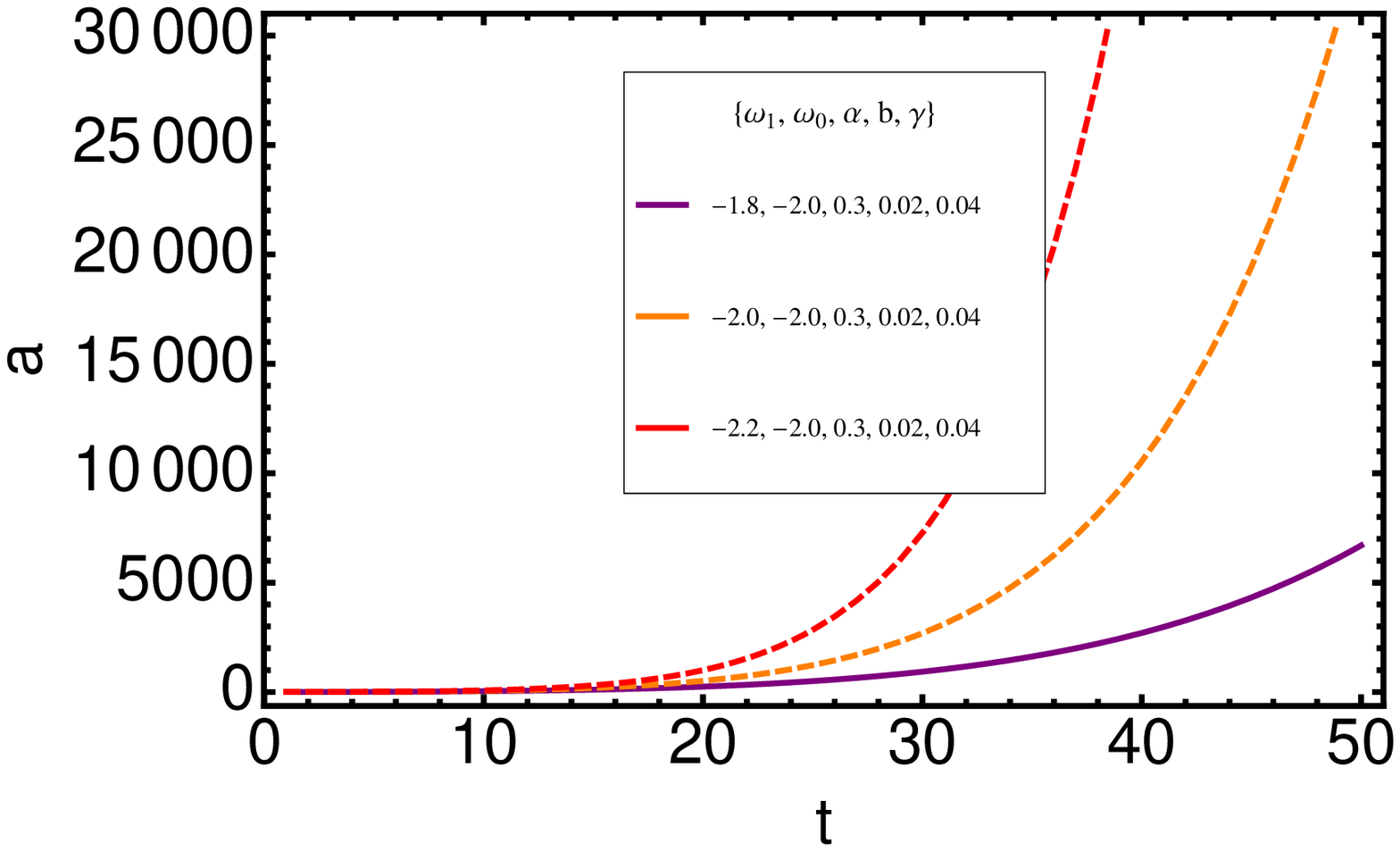}&
\includegraphics[width=50 mm]{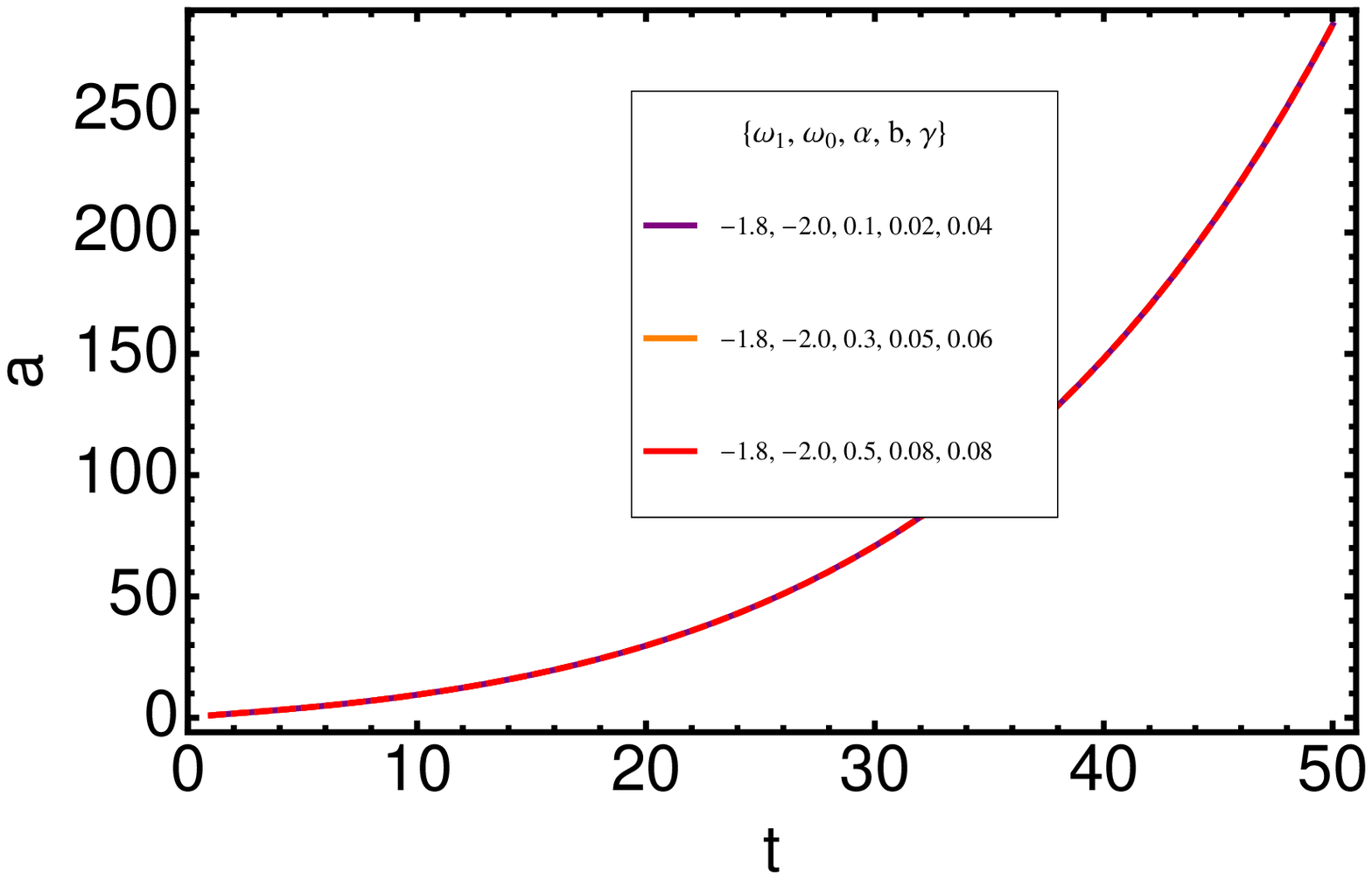}
 \end{array}$
 \end{center}
\caption{Scale factor of two component fluid Universe}
 \label{fig:10}
\end{figure}

\begin{figure}[h]
 \begin{center}$
 \begin{array}{cccc}
\includegraphics[width=50 mm]{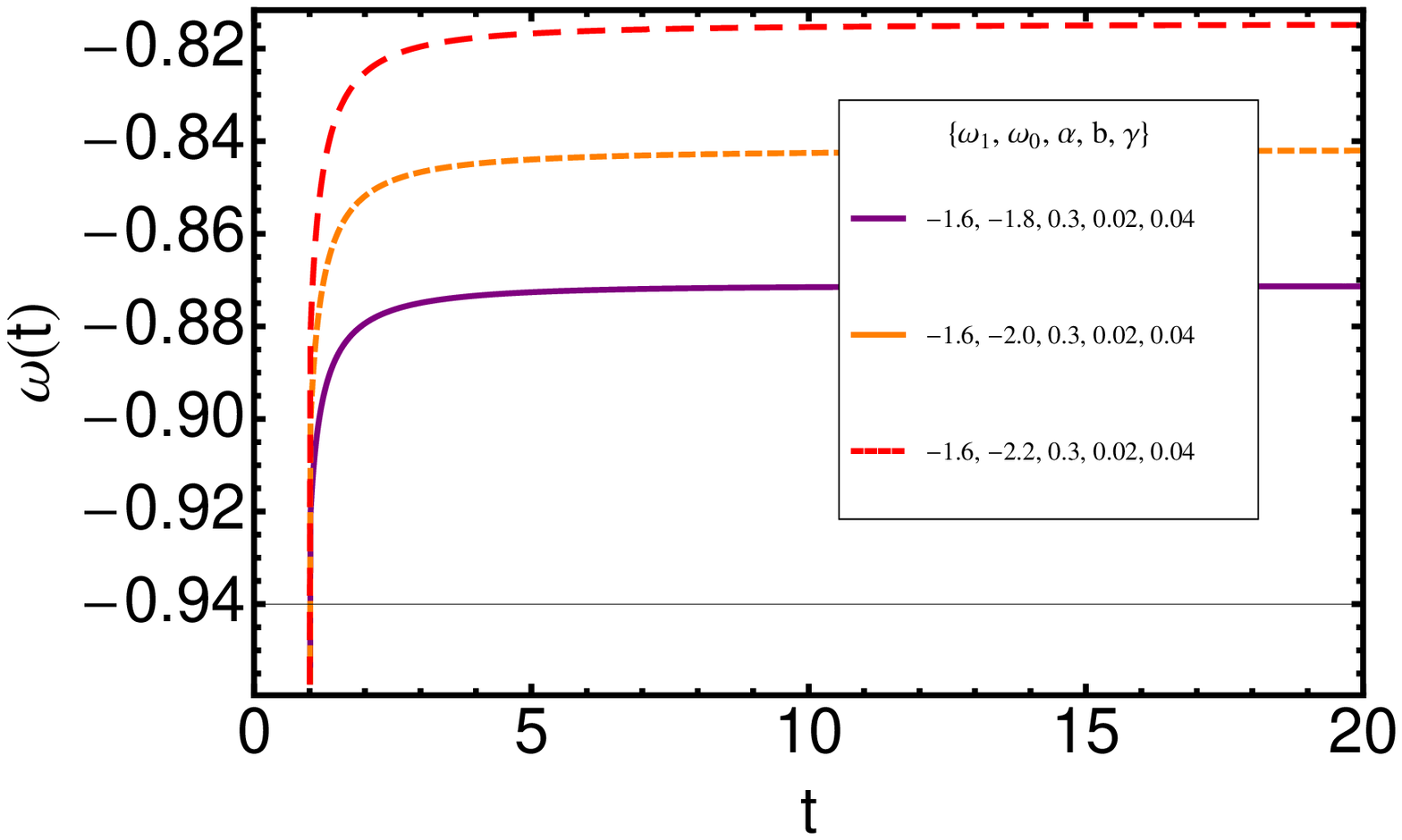} &
\includegraphics[width=50 mm]{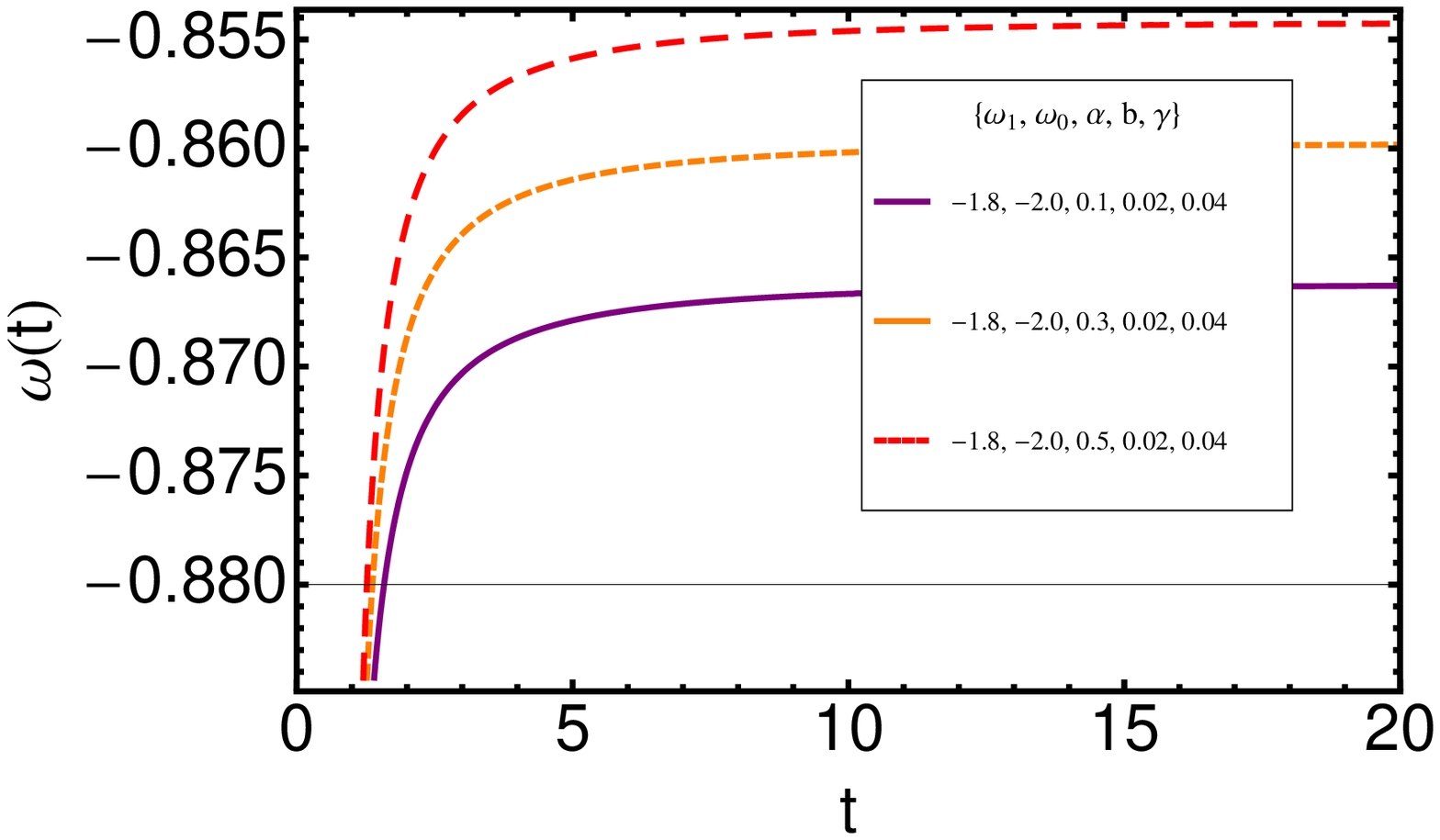}\\
\includegraphics[width=50 mm]{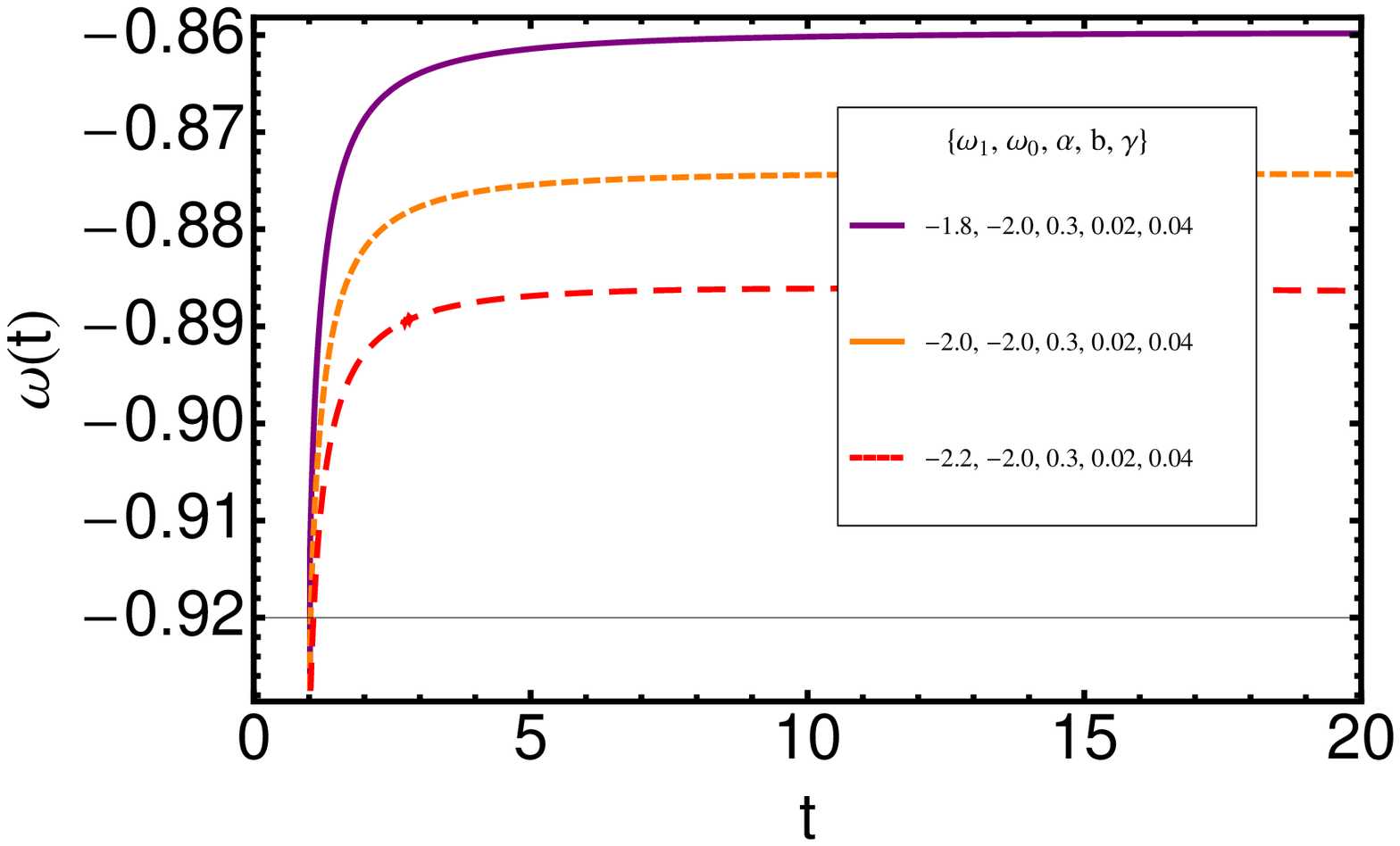} &
\includegraphics[width=50 mm]{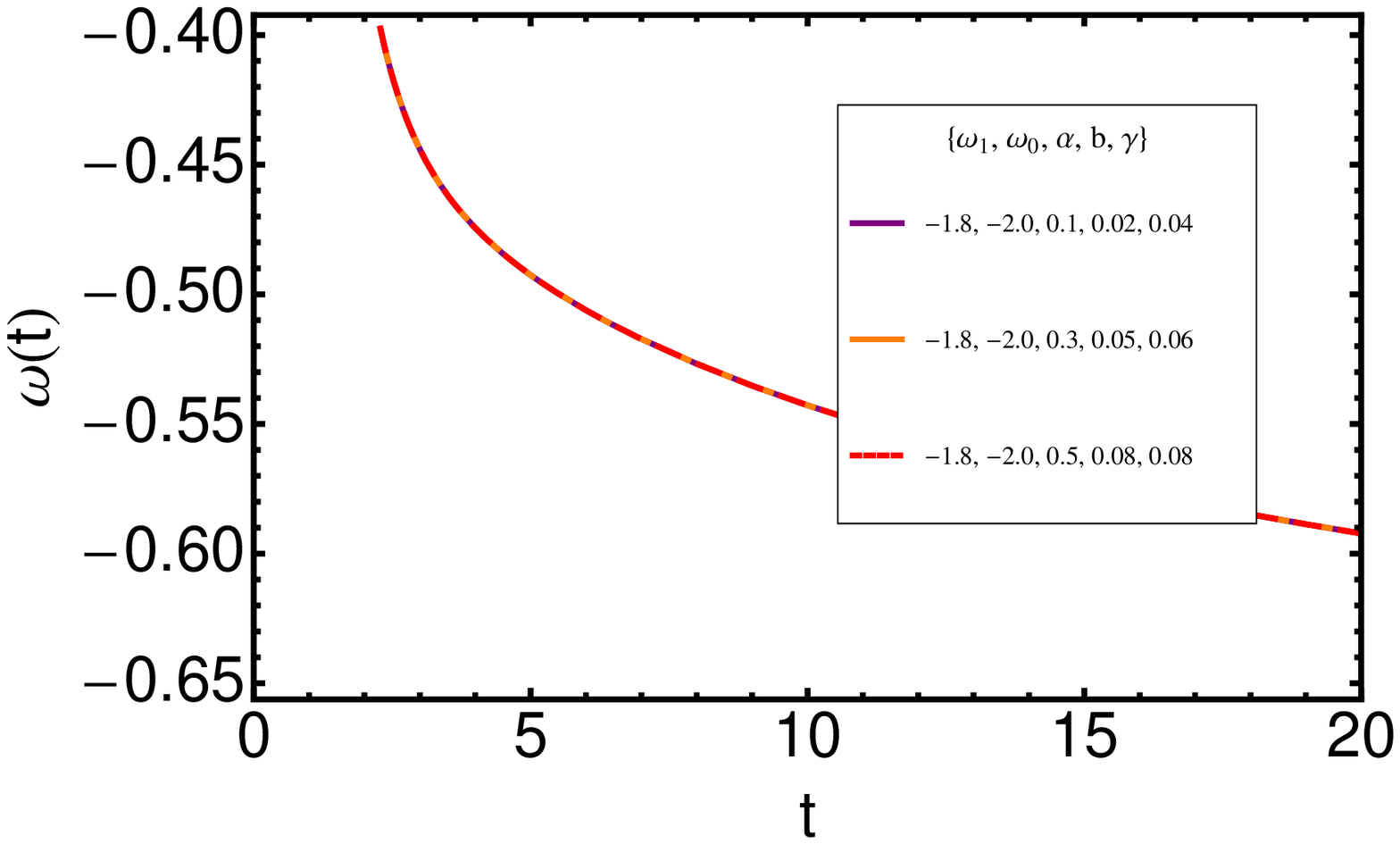}
 \end{array}$
 \end{center}
\caption{EoS parameter of two component fluid Universe}
 \label{fig:11}
\end{figure}

\begin{figure}[h]
 \begin{center}$
 \begin{array}{cccc}
 \includegraphics[width=60 mm]{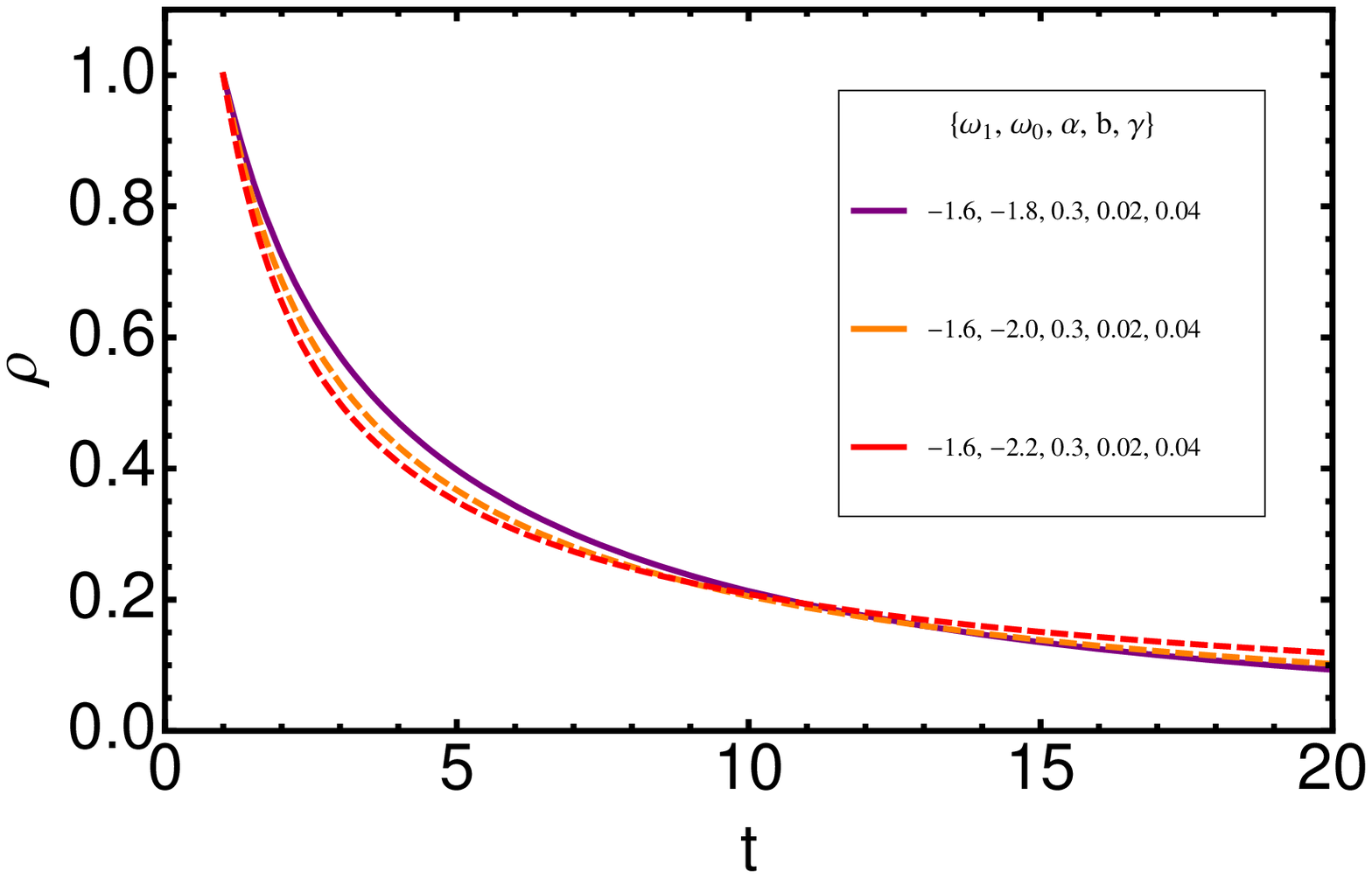} &
 \includegraphics[width=60 mm]{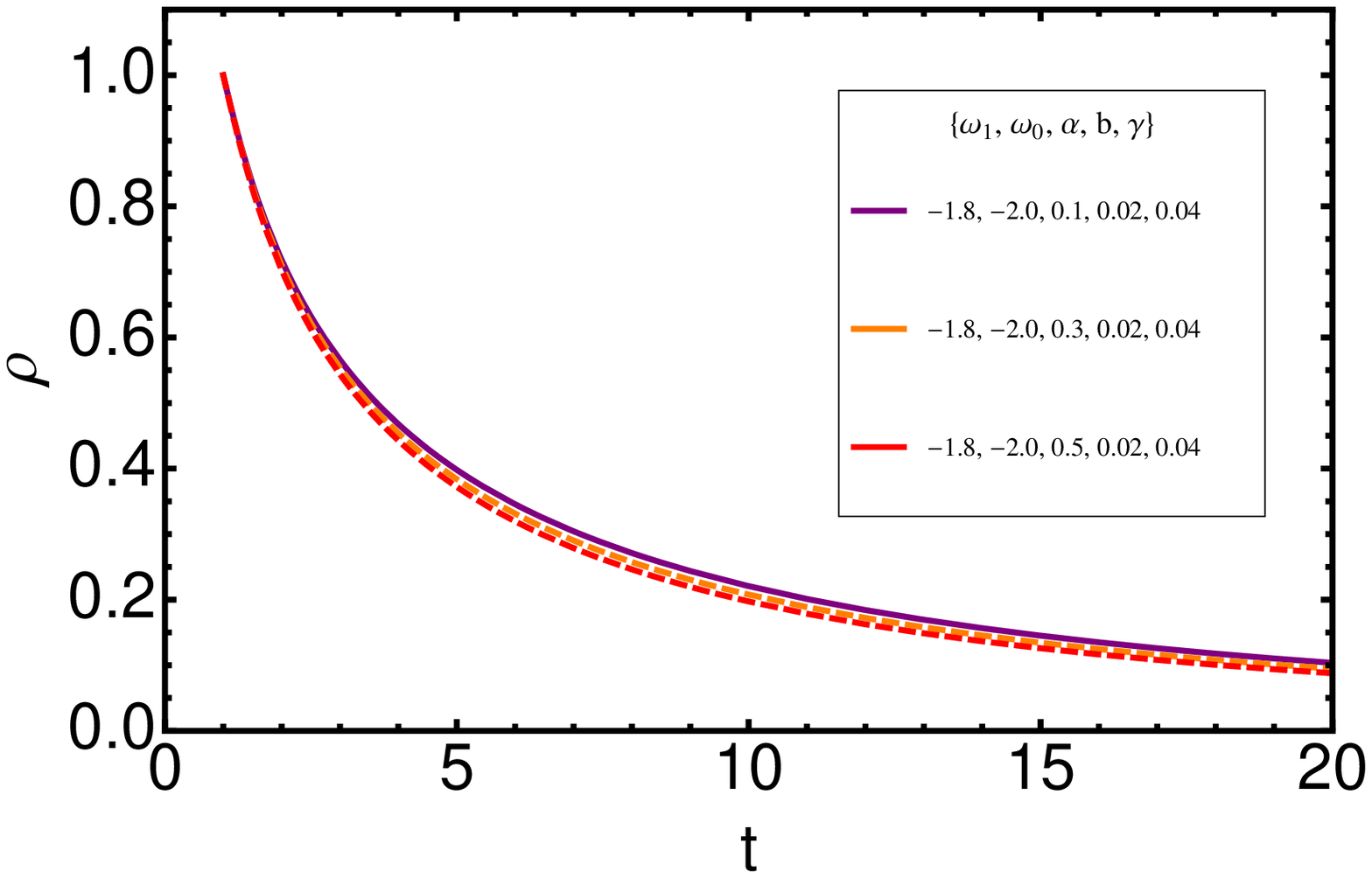} \\
 \includegraphics[width=60 mm]{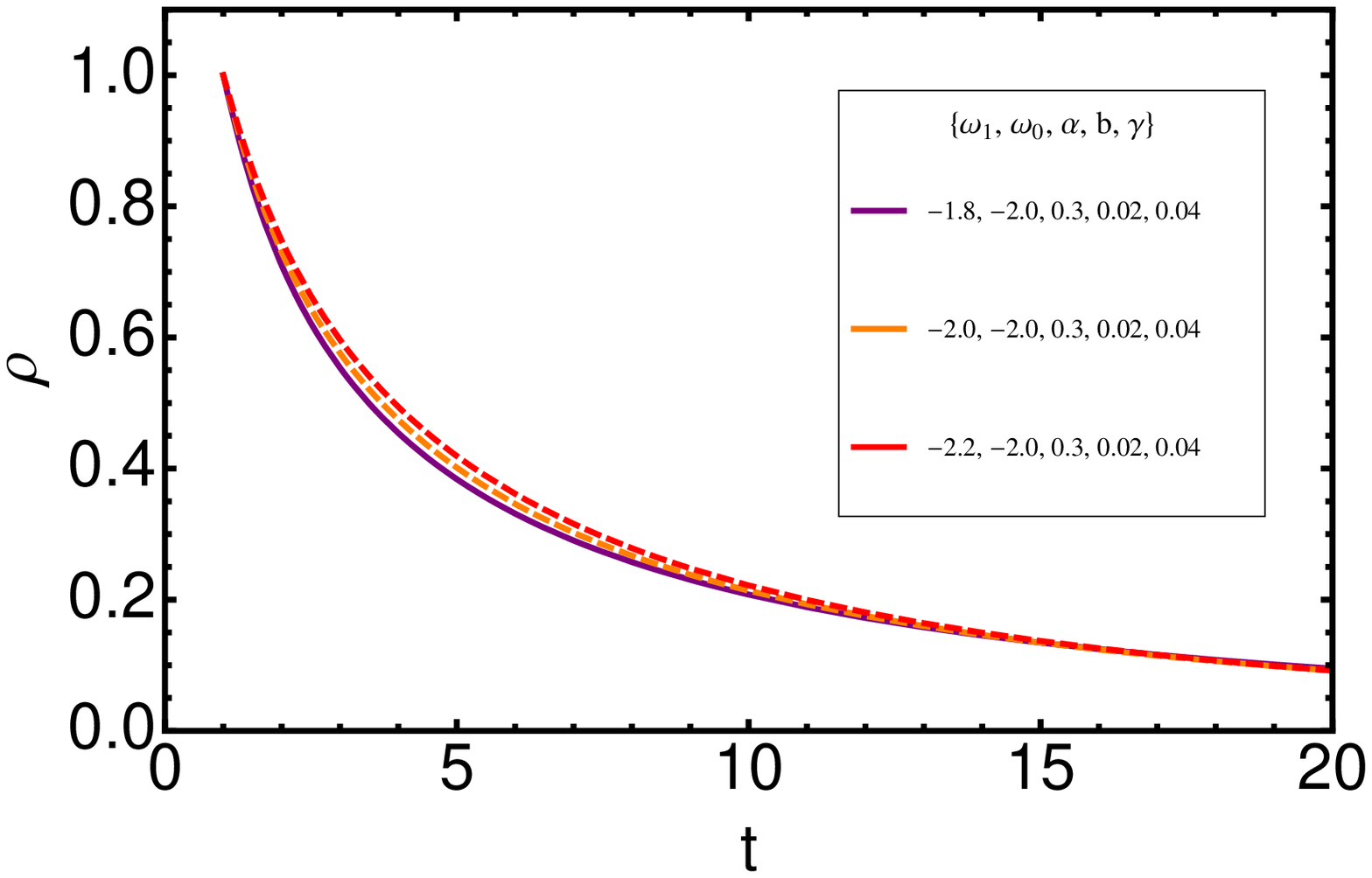} &
 \includegraphics[width=60 mm]{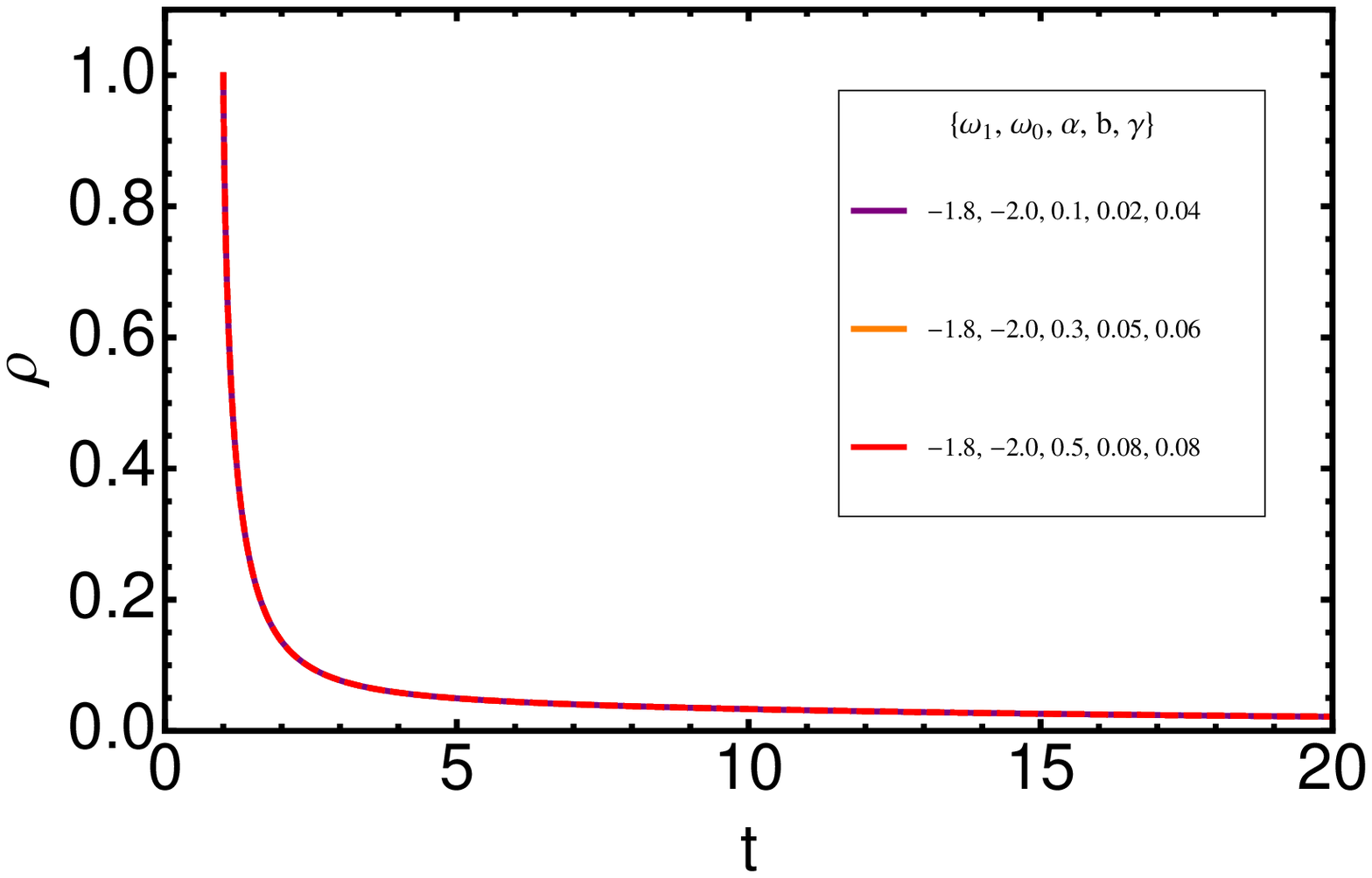}
 \end{array}$
 \end{center}
 \caption{Energy density of two component fluid Universe}
 \label{fig:12}
\end{figure}

\end{document}